%% file: main.tex
\documentclass[11pt]{article}
\usepackage{moreverb}
\usepackage[utf8]{inputenc} 
\usepackage{hyperref}
 
\usepackage{amsmath}
\usepackage{enumitem}
\usepackage{subcaption}

\usepackage{lscape}
\usepackage{adjustbox}

\usepackage[legalpaper, margin=1in]{geometry}

\usepackage{tikz}
\usetikzlibrary{positioning}

\usepackage{todonotes}

\usepackage{empheq}

\usepackage{listings}
\definecolor{codegreen}{rgb}{0,0.6,0}
\definecolor{codegray}{rgb}{0.5,0.5,0.5}
\definecolor{codepurple}{rgb}{0.58,0,0.82}
\definecolor{backcolour}{rgb}{0.99,0.99,0.97}
\lstdefinestyle{stan}{
	literate={~}{$\sim$}{1},
	backgroundcolor=\color{backcolour},   
	commentstyle=\color{codegreen},
	keywords = {real, vector, matrix, model, parameters, transformed, target, int, return, function},
	keywordstyle=\color{magenta},
	numberstyle=\tiny\color{codegray},
	stringstyle=\color{codepurple},
	emph={%
		normal, cauchy, inv_gamma, bernoulli_logit, gamma, laplace_marginal_*_lpmf %
	},
	emphstyle=\color{codepurple},%
	basicstyle={\ttfamily},
	breaklines=true,                 
	keepspaces=true,                 
	showspaces=false,                
}
\title{Bayesian workflow for disease transmission modeling in Stan}

\author{Léo Grinsztajn\footnote{École polytechnique, Palaiseau, France, \url{leo.grinsztajn@polytechnique.edu}}, 
            Elizaveta Semenova\footnote{Data Sciences and Quantitative Biology, Discovery Sciences, R\&D, AstraZeneca, Cambridge, UK},
            Charles C.~Margossian\footnote{Department of Statistics, Columbia University, New York, NY, USA}, and
            Julien Riou\footnote{Institute of Social and Preventive Medicine, University of Bern, Bern, Switzerland}}

\date{}

\begin{document}

\maketitle

\begin{abstract}
	This tutorial shows how to build, fit, and criticize disease transmission models in Stan, and should be useful to researchers interested in modeling the SARS-CoV-2 pandemic and other infectious diseases in a Bayesian framework.
	Bayesian modeling provides a principled way to quantify uncertainty and incorporate both data and prior knowledge into the model estimates.  
	Stan is an expressive probabilistic programming language that abstracts the inference and allows users to focus on the modeling. 
	As a result, Stan code is readable and easily extensible, which makes the modeler’s work more transparent. 
	Furthermore, Stan’s main inference engine, Hamiltonian Monte Carlo sampling, is amiable to diagnostics, which means the user can verify whether the obtained inference is reliable.
	In this tutorial, we demonstrate how to formulate, fit, and diagnose a compartmental transmission model in Stan, first with a simple Susceptible-Infected-Recovered (SIR) model, then with a more elaborate transmission model used during the SARS-CoV-2 pandemic. 
	We also cover advanced topics which can further help practitioners fit sophisticated models; notably, how to use simulations to probe the model and priors, and computational techniques to scale-up models based on ordinary differential equations. 
\end{abstract}

\section{Introduction}\label{sec:intro}
\input{intro.tex}

\section{Bayesian modeling in Stan}\label{sec:bayes}
\input{bayes_intro.tex}

\section{Bayesian modeling workflow}\label{sec:workflow}

\input{workflow.tex}

\section{Simple SIR model}\label{sec:simplesir}
\input{simple_sir_no_code.tex}

%
%
\section{Scaling-up ODE-based models}\label{sec:scaleup}
\input{scaleup.tex}

%
\section{Application to SARS-CoV-2 transmission in Switzerland}\label{sec:extension}
\input{extension.tex}

\section{Conclusions}\label{sec:conclusion}
\input{conclusion.tex}
%


\section*{Acknowledgments}

We thank Ben Bales and Andrew Gelman for their helpful comments.

\section*{Author contributions}

LG, ES, CCM and JR conceived this tutorial and wrote the manuscript. LG and ES performed the computations.

\section*{Financial disclosure}

ES was supported by AstraZeneca postdoc programme.
JR is funded by the Swiss National Science Foundation (grant 174281).

\section*{Conflict of interest}

The authors declare no conflict of interests.

\clearpage

\bibliography{tuto_bibliography}
\bibliographystyle{unsrt}

\end{document}

%% file: intro.tex
The pandemic of severe acute respiratory syndrome coronavirus 2 (SARS-CoV-2) has led to a renewed interest in infectious disease modeling and, amongst other approaches, Bayesian modeling. 
%
%
Well-constructed models allow researchers to infer the value of key epidemiological parameters required to inform public health policies.
During the early stages of the pandemic, modelers have assessed the effect of control interventions on transmission \cite{flaxman2020estimating}, quantified the burden of the epidemic \cite{salje2020covid}, and estimated the mortality rate after adjusting for reporting biases\cite{riou2020covid}, to only name a few examples.
%

Mechanistic disease transmission models mirror natural phenomena such as contagion, incubation and immunity \cite{keeling2009mathematical}.
These models can operate at different scales, the most important being the individual scale (individual or agent-based models) and the population scale (population-based or compartmental models). 
Agent-based models are used to simulate the occurrence of stochastic events, such as transmission or symptom onset, in time for every individual of the population.
These models can reach a high level of sophistication but are computationally expensive.
Agent-based models are also notoriously difficult to parametrize.

Population-based or compartmental models subdivide the total population into homogeneous groups, called compartments.
Individuals within a compartment are considered to be in the same state with regards to the natural history of the disease.
These states may for example be ``susceptible'', ``infectious'', and ``recovered''.
We can model transitions between compartments as deterministic or stochastic.
In the deterministic framework, the flows between compartments can be described by a system of ordinary differential equations (ODEs). 
This implies that the size of each compartment at every time point can be obtained by numerically solving the system of ODEs, with the same parametrization always 
leading to the same outcome.
The flows between compartments can also be simulated stochastically, leading to the same results on average over multiple simulations (ignoring disease extinctions) but with a better handling of uncertainty that comes at a generally higher computational price.
This makes stochastic compartmental models more adapted to low-level transmission and small populations. 
In other situations, deterministic compartmental models are easier to formulate and computationally tractable, which makes them more adapted to tasks that require simulating the system a large number of times, such as model fitting.
This article focuses on deterministic compartmental models.

\texttt{Stan} is a probabilistic programming framework primarily used for Bayesian inference and designed to let the user focus on modeling, while inference occurs under the hood \cite{carpenter2017stan}.
One goal of the \texttt{Stan} project is not merely to improve model fitting but to make model development more efficient.
This development process includes building, debugging, improving, and expanding models as more data and knowledge become available, and motivates the concept of the \textit{Bayesian modeling workflow}
\cite{gabry2019visualization, betancourt2020principled,gelman2020bayesian}, a central topic in this article.
\texttt{Stan} is an expressive language that supports many probability densities, matrix operations, and numerical ODE solvers.
We can combine these elements to specify a broad range of data generating processes.
Generative models formulated in \texttt{Stan} can be used both for simulation-based prediction and for parameter inference. 
In the context of epidemiological modeling, the powerful framework can help us estimate such crucial parameters as the basic reproduction number $\mathcal{R}_0$ or the infection-fatality ratio from observed data.
\texttt{Stan} bolsters several inference methods: full Bayesian inference using Markov Chain Monte Carlo (MCMC), approximate Bayesian inference with variational inference, and penalized maximum likelihood estimation. 
We focus here on Bayesian inference with MCMC, specifically with the dynamic Hamiltonian Monte Carlo (HMC) sampler \cite{hoffman2014nuts, betancourt2018conceptual}.
Bayesian inference gives us a principled quantification of uncertainty and the ability to incorporate domain knowledge in the form of priors, while HMC is a reliable and flexible algorithm. 
In addition, \texttt{Stan} provides diagnostic tools to evaluate both the reliability of the inference 
and the adequacy of the model.

This tutorial examines how to formulate, fit, and diagnose compartmental models for disease transmission in Stan.
We first focus on a simple Susceptible-Infected-Recovered (SIR) model, before tackling a more sophisticated model of SARS-CoV-2 transmission.
Rather than only show the results from a polished model, we devote much attention to dealing with flawed models and flawed inference; and we explore how different modes of failure can help us develop improved models and better tune our inference algorithms.
Naturally, \texttt{Stan} is not the only tool available to practitioners and many of the concepts we discuss can be deployed using other probabilistic programming languages.
At times, we dig into the specific mechanics of \texttt{Stan} in order to provide a more practical discussion on modeling and computational efficiency.
%
%
A complementary notebook\cite{grinsztajn2020bayesian} with the full code, and a Github repository\footnote{\url{https://github.com/charlesm93/disease_transmission_workflow}}containing the relevant scripts are available online.
Throughout the tutorial, we use \texttt{R} as a scripting language (\texttt{Stan} can also be used with other languages such as \texttt{Python}, \texttt{Julia} or \texttt{Matlab}).
While we review some elementary concepts, we assume that the reader has basic familiarity with Bayesian inference and Stan. Other tutorials on the subject include the work by Chatzilena et al.\cite{chatzilena2019contemporary} and Mihaljevic \cite{mihaljevic2016tutorial} on transmission models, and the case studies by Carpenter \cite{carpenter2018predator-prey}, Weber \cite{weber2018ode}, and Margossian and Gillespie \cite{margossian2017ode} on ODE-based models, all of which can serve as complementary reading.

%

%% file: bayes_intro.tex
\texttt{Stan} is a tool which provides both a language to formulate probabilistic models and methods to do inference on these models. Before using \texttt{Stan} to model disease transmission, we quickly review how to specify a probabilistic model. A more thorough introduction to the topic can be found here \cite{gelman2013bda}.

\subsection{Specifying a model}

We can specify a Bayesian model by defining a joint distribution over observed variables, $\mathcal Y$, and  unobserved variables, $\theta$,
$$ p(\mathcal Y, \theta).$$
In this tutorial, $\theta$ is simply the set of unknown model parameters.
A \texttt{Stan} file defines a procedure to evaluate the log density, $\log p(\mathcal Y, \theta)$.
This joint distribution conveniently decomposes into two terms - the \textit{prior density} $p(\theta)$ and the \textit{sampling density} (or \textit{likelihood}) $p(\mathcal Y \mid \theta)$, defining a generating process for  $\mathcal Y$ given parameters $\theta$  :
$$
  p(\mathcal Y, \theta) = p(\theta) p(\mathcal Y \mid \theta).
$$
%

\subsection{Bayesian inference}

Inference reverse-engineers the data generating process and aims to estimate parameter values given the observations. 
In a Bayesian framework, the set of plausible parameter values conditional on the data is characterized by the \textit{posterior distribution}. The posterior distribution combines information from the data and prior knowledge and is obtained via Bayes' rule 
$$
  p(\theta \mid \mathcal Y) \propto p(\theta) p(\mathcal Y \mid \theta).
$$
An analytical expression for $p(\theta \mid \mathcal Y)$ is rarely available and we must rely on inference algorithms to learn about the posterior distribution.
One general strategy is to draw approximate samples from the posterior distribution and use these to construct sample estimates of the posterior mean, variance, median, quantiles, and other quantities of interest.
This leads to a general class of algorithms called Markov chain Monte Carlo (MCMC) samplers.

\paragraph{Hamiltonian Monte Carlo.} \texttt{Stan} supports dynamic Hamiltonian Monte Carlo \cite{betancourt2018conceptual, radford2012hmc} (HMC), a widely popular MCMC method.
HMC exploits the gradient of $\log p(\theta, \mathcal Y)$ to simulate trajectories, with acceleration informed by the local geometry of the posterior density.
This leads to a rapid exploration of the parameter space and reduces correlation between successive samples.
HMC has been shown to scale better than random walk samplers, e.g. Metropolis and Gibbs, when exploring high-dimensional spaces or when parameters exhibit a strong posterior correlation\cite{hoffman2014nuts}.
The method has been successfully applied across a broad range of problems.
When HMC fails, it typically does so ``loudly'', meaning diagnostics can reliably determine whether or not the inference should be trusted.
While the method was first proposed in 1987\cite{duane1987hmc}, its challenging implementation prevented its wide adoption by the statistics and scientific community.
Indeed, calculating the gradient of $\log p(\theta, \mathcal Y)$ is a cumbersome task and the original HMC algorithm entails many tuning parameters which, without proper tuning, result in suboptimal performance.
The advent of automatic differentiation\cite{baydin2018ad, margossian2019ad} and of the No-U-Turn sampler, an adaptive HMC algorithm\cite{hoffman2014nuts}, in large parts resolved these problems, making it straightforward to apply the algorithm across a broad range of models.
With a software such as \texttt{Stan}, it is indeed possible to revise a model without needing to rewrite the HMC sampler.
In summary, HMC works well in high-dimensional spaces, can handle the intricate posterior geometry that arises in sophisticated models, can be readily applied to several models, and is amiable to diagnostics.
For these reasons, we find HMC to be very well suited for a Bayesian workflow (Section~\ref{sec:workflow}).
Naturally, the algorithm is not without limitation: for example, HMC cannot be applied to discrete problems without marginalization or continuous relaxation. Furthermore, densities with difficult geometries (e.g. multiple modes, heavy tails, varying scales) can frustrate the algorithm\cite{stan2020problematic}\cite{stan2020reparametrization}
and, while being broadly applicable, HMC can be slower than specialized algorithms used for certain statistical models.

\subsection{Getting started with Stan}


\paragraph*{Installation}
At its core, \texttt{Stan} is written in \texttt{C++} but for convenience it can be interfaced with a scripting language, such as \texttt{R} and \texttt{Python}; see \href{https://mc-stan.org/users/interfaces/}{the Stan interface page}\cite{stan2020interfaces} for instructions on how to install these interfaces.
This tutorial uses \texttt{RStan} and \texttt{R}.

\paragraph*{Learning Stan}

While we do not assume the reader is already familiar with \texttt{Stan}, we encourage them to consult one of the many resources available to learn the language.
As a starting point, we recommend the introduction by Betancourt\cite{betancourt2020introduction}.
The \href{https://mc-stan.org/docs/2_26/reference-manual/index.html}{reference manual}\cite{stan2020reference} documents the language's syntax;
the \href{https://mc-stan.org/docs/2_26/stan-users-guide/index.html}{Stan user guide}\cite{stan2020user} provides the code for numerous models.
Guidelines on how to run \texttt{Stan} and various diagnostics are available on the \href{https://mc-stan.org/users/interfaces/}{Stan interface page}\cite{stan2020interfaces}.
A complementary notebook\cite{grinsztajn2020bayesian} offers a detailed walk-through of the example in Section~\ref{sec:simplesir} and can be used as a hands-on way to learn the language.


\paragraph*{Going forward}
To learn more, the reader may find it helpful to read some of the \href{https://mc-stan.org/users/documentation/case-studies.html}{cases studies}\cite{stan2020case} and \href{https://mc-stan.org/users/documentation/tutorials.html}{tutorials}\cite{stan2020tutorial} on specific applications of \texttt{Stan}. We also recommend the \href{https://discourse.mc-stan.org/}{Stan forum} as a place where modelers can discuss their models with other researchers.

\subsection{Coding a model in Stan} \label{sec:blocks}

\begin{figure}[!htbp]
  \begin{center}
    \begin{tikzpicture}
  [
    Box/.style={rectangle, draw=black!, fill=green!0, thick, minimum size=10mm},
    Gray/.style={rectangle, draw=black!, fill=gray!35, thick, minimum size=1mm},
    Round/.style={circle, draw=black!, fill=green!0, thick, minimum size=1mm},
  ]
  \node[Box, text width=100mm] (Data) at(0, -.2) {{\large \texttt{data}} \vspace{1.5mm} \newline Declare known variables, which remain fixed while running MCMC. This includes covariates, observed responses, and other constants. \newline \newline \textit{No operations to perform.}
};
    \node[Box, text width=100mm] (TransformedD) at(0, -3.5) {{\large \texttt{transformed data}} \vspace{1.5mm} \newline Declare additional fixed variables and do operations on variables declared in \texttt{data} and in this block. This is useful when we want to work with transformations of our original data or create constants for bookkeeping purposes.
    \newline \newline  \textit{Operations are performed once per model fit.}
};
  \node[Box, text width=100mm] (Parameters) at(0, -7) {{\large \texttt{parameters}} \vspace{1.5mm} \newline Declare unknown variables, typically model parameters. When running MCMC this also defines the space over which we run the Markov chain.
  \newline \newline \textit{No operations to perform.}
  };
    \node[Box, text width=100mm] (Parameters) at(0, -10.3) {{\large \texttt{transformed parameters}} \vspace{1.5mm} \newline Run operations on the parameters. This is, for instance, where we will solve differential equations which depend on model parameters.
    \newline \newline \textit{Operations are performed and differentiated once per integration step, i.e. multiple times per iteration.}
    };
    \node[Box, text width=100mm] (Parameters) at(0, -13.7) {{\large \texttt{model}} \vspace{1.5mm} \newline Compute $\log p(\mathcal Y, \theta)$, using operations on variables declared in the previous blocks.
    \newline \newline \textit{Operations are performed and differentiated once per integration step, i.e. multiple times per iteration.}
    };
    \node[Box, text width=100mm] (Parameters) at(0, -17.5) {{\large \texttt{generated quantities}} \vspace{1.5mm} \newline Run operations on all previously declared variables to compute quantities of interest, such as derived variables and predictions. The variables we compute here depend on the parameters, but crucially none of the operations in this block contribute to computing $\log p(\mathcal Y, \theta)$.
     \newline \newline \textit{Operations are performed once per iteration.}};

  \path[->, draw, thick] (-6, 0.5) -- (-6, -19); 
  \end{tikzpicture}
  \end{center}
  \caption{Coding blocks in a \texttt{Stan} file. The operations in certain blocks are performed multiple times and in some cases differentiated; as a result, the computational cost of fitting the model is dominated by the transformed parameters and model blocks. Not shown is the \texttt{functions} block, which defines functions that can be called in any of the operative blocks.}
  \label{fig:blocks}
\end{figure}
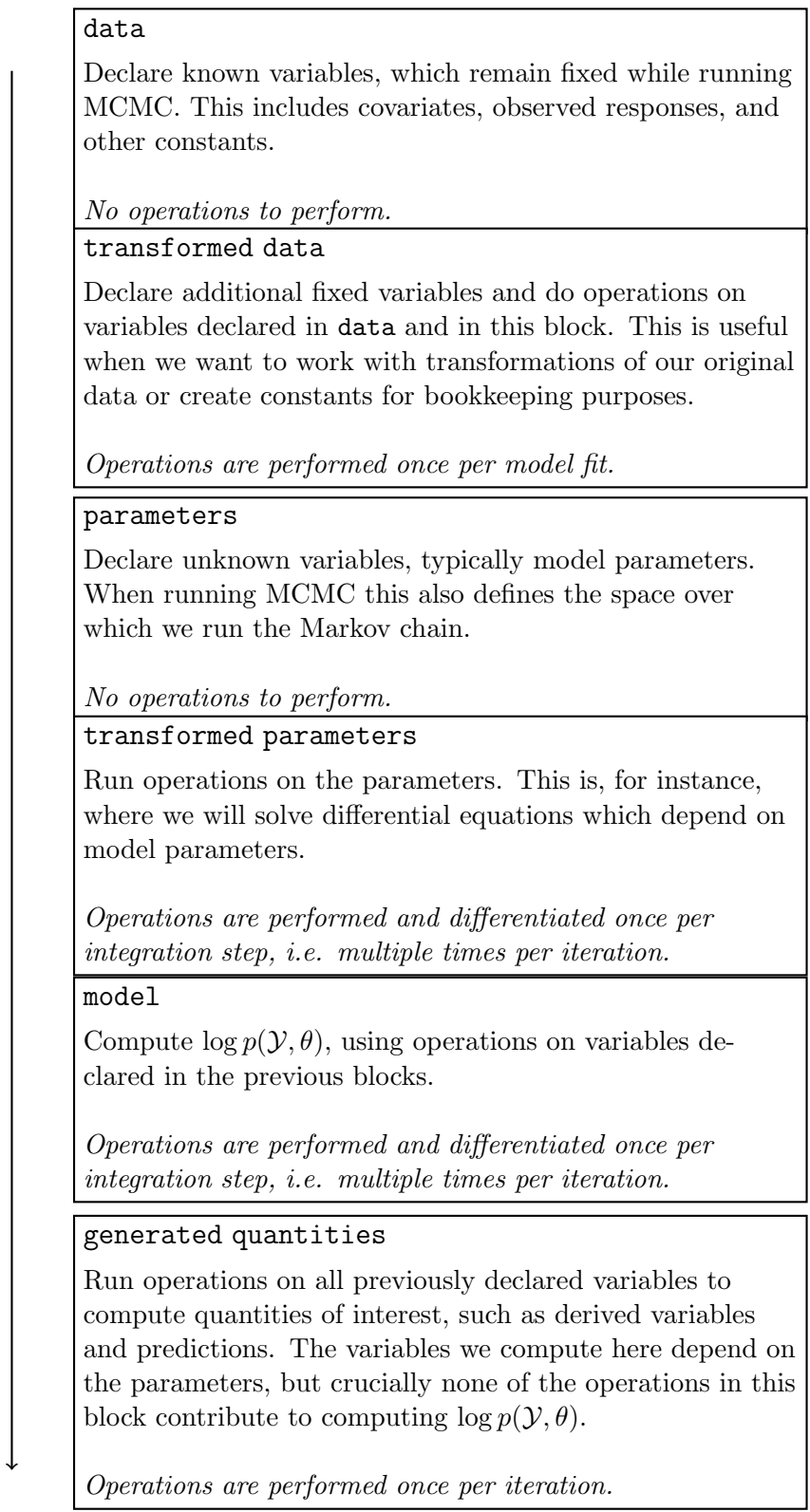

Using a \texttt{Stan} file, we can specify a model in a way that is suitable to Bayesian inference.
This entails (i) defining a procedure to compute the log joint density, $\log p(\mathcal Y, \theta)$, and (ii) identifying which variables are fixed data, $\mathcal Y$, and which ones are latent variables, $\theta$.
For pedagogical and computational reasons, a \texttt{Stan} file is divided into \textit{coding blocks}, which help us distinguish the different components of our model.
Figure~\ref{fig:blocks} provides a description of each block and indicates the order in which the blocks are executed. The three main blocks are \texttt{data}, \texttt{parameters}, and \texttt{model}.
In the \texttt{data} block, we declare the fixed variables, $\mathcal Y$, and in the \texttt{parameters} block the model parameters, $\theta$.
In the \texttt{model} block, the user specifies a set of operations on the declared variables to compute the log joint density, $\log p(\mathcal Y, \theta)$.
It is possible to do further operations on the fixed data and on the model parameters by using the \texttt{transformed data} and the \texttt{transformed parameters} blocks, respectively.
Additional quantities which depend on the parameters, but are not required to compute the log joint density -- for example, predictions -- can be computed in the \texttt{generated quantities} block.
Finally, the \texttt{function} block can be used to declare functions which we can call in all the operative blocks, i.e. all blocks but \texttt{data} and \texttt{parameters}.

Section~\ref{sec:scaleup} offers more details on \texttt{Stan}'s inference engine and discusses how each block interacts with the inference algorithm.
For now, we simply note that the computational cost of fitting a model is dominated by operations in the \texttt{transformed parameters} and \texttt{model} blocks.

%% file: workflow.tex
\begin{figure}[htbp]
	\begin{center}
		\scalebox{.9}{ \begin{tikzpicture}
				[
				Box/.style={rectangle, draw=black!, minimum size=10mm, text width=25mm,text centered},
				Box2/.style={rectangle, draw=black!, fill=gray!20, minimum size=10mm, text width=20mm,text centered,rounded corners},
				Round/.style={circle, draw=black!, fill=green!0, minimum size=1mm},
				]
				\node[Round] (Data) at(4, -4) {Data};
				\node[Box] (Model) at(-4, 0) {Build model};
				\node[Box2] (Crit1) at (0, 0) {Criticize model};
				\node[Box] (Infer) at (0, -2) {Fit model};
				\node[Box2] (Crit2) at (4, -2) {Criticize inference};
				\node[Box2] (Crit3) at (8, -2) {Criticize fitted model};
				\node[Box] (App) at (8,-4) {Apply fitted model};
				
				\path [->, draw] (Model) -- (Crit1);
				\path [->, draw] (Crit1) -- (Infer);
				\path [->, draw] (Infer) -- (Crit2);
				\path [->, draw] (Crit2) -- (Crit3);
				\path [->, draw] (Crit3) -- (App);
				\path [->, draw] (Crit1) edge[bend right=20] (Model);
				\path [->, draw] (Crit2) edge[bend right=40] (Infer);
				\path [->, draw] (Crit2) edge[bend right=40] (Model);
				\path [->, draw] (Crit3) edge[bend right=42] (Model);
				\path [->, draw, dotted] (Data) -- (Infer);
				\path [->, draw, dotted] (Data) -- (Crit2);
				\path [->, draw, dotted] (Data) -- (Crit3);
		\end{tikzpicture}}
	\end{center}
	\caption{Model development as an iterative process.}
	\label{fig:loop}
\end{figure}
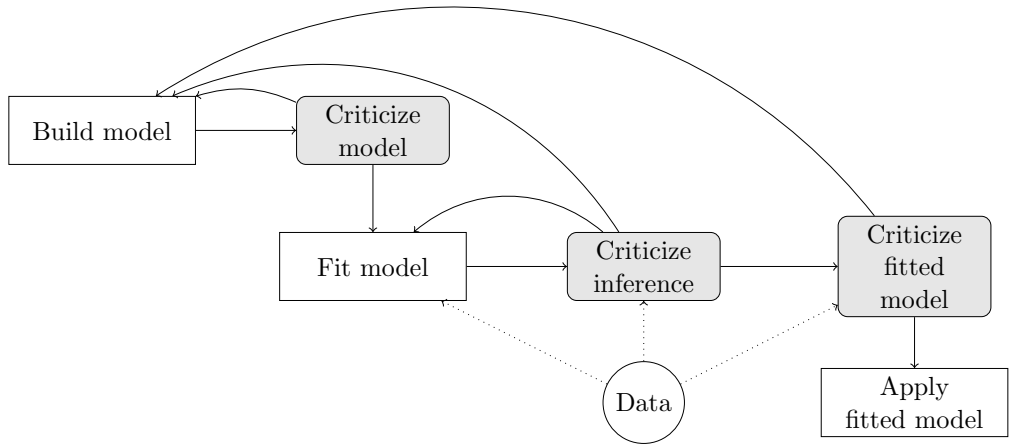

The notion of a modeling workflow has likely existed in one form or the other for quite some time.
One useful illustration of this concept is Box's loop \cite{box1976science, blei2014build}. 
The loop prescribes the following workflow: build a model, fit the model, criticize, and repeat.
We find it useful to distinguish three parts in the criticism step: (i) troubleshoot the model before fitting it,
(ii) criticize the inference after attempting a fit,
and (iii) once the inference is deemed reliable, criticize the fitted model (Figure~\ref{fig:loop}).
The goal of the criticism is to identify shortcomings in our methods and motivate adjustments,
which is why the process loops back to the first and second steps.
%
%
%

The first takeaway from this concept is that model development is an iterative process:
we do not expect the first model we devise to be perfect and are prepared to make adjustments.
Potential limitations in our model can include implementation errors, inappropriate priors or failure to account for a crucial step in the data generating process.
It is essential to detect and correct these problems before we apply the model.
Several points can make this potentially tedious process easier. 
First, our probabilistic programming language must be expressive enough to accommodate our initial model and subsequent revisions.
Second, we prefer efficient and flexible inference algorithms that can operate quickly and do not require tuning each time we revise the model.
Third, we require the inference to be reliable, meaning we can diagnose many different types of failures, and hopefully adjust our algorithms or even our model to overcome issues we identify.
This is especially important when we care about the posterior distribution of our parameters, as we might in epidemiological models where the parameters have a scientific interpretation. We must realise, however, that even when the inference is reliable, the model may fail to solve the scientific problem at hand.

\texttt{Stan} and its HMC sampler typically meet these criteria, and provide solutions and features adapted to each of the goals outlined in our modeling workflow.
Predictions computed across prior distributions (prior predictive checks) are useful to detect implementation errors and assess the adequacy of the chosen priors for the problem at hand \cite{gabry2019visualization}.
\texttt{Stan}'s HMC sampler provides multiple inference diagnostics, including the detection of bias in sampling by running multiple MCMC chains and monitoring mixing, energy and divergent transitions.
Finally, we can study the implications of the fitted model by computing predictions across the posterior distribution (posterior predictive checks).

%


%% file: simple_sir_no_code.tex
\begin{figure}
    \centering
    \includegraphics[width=.5\linewidth]{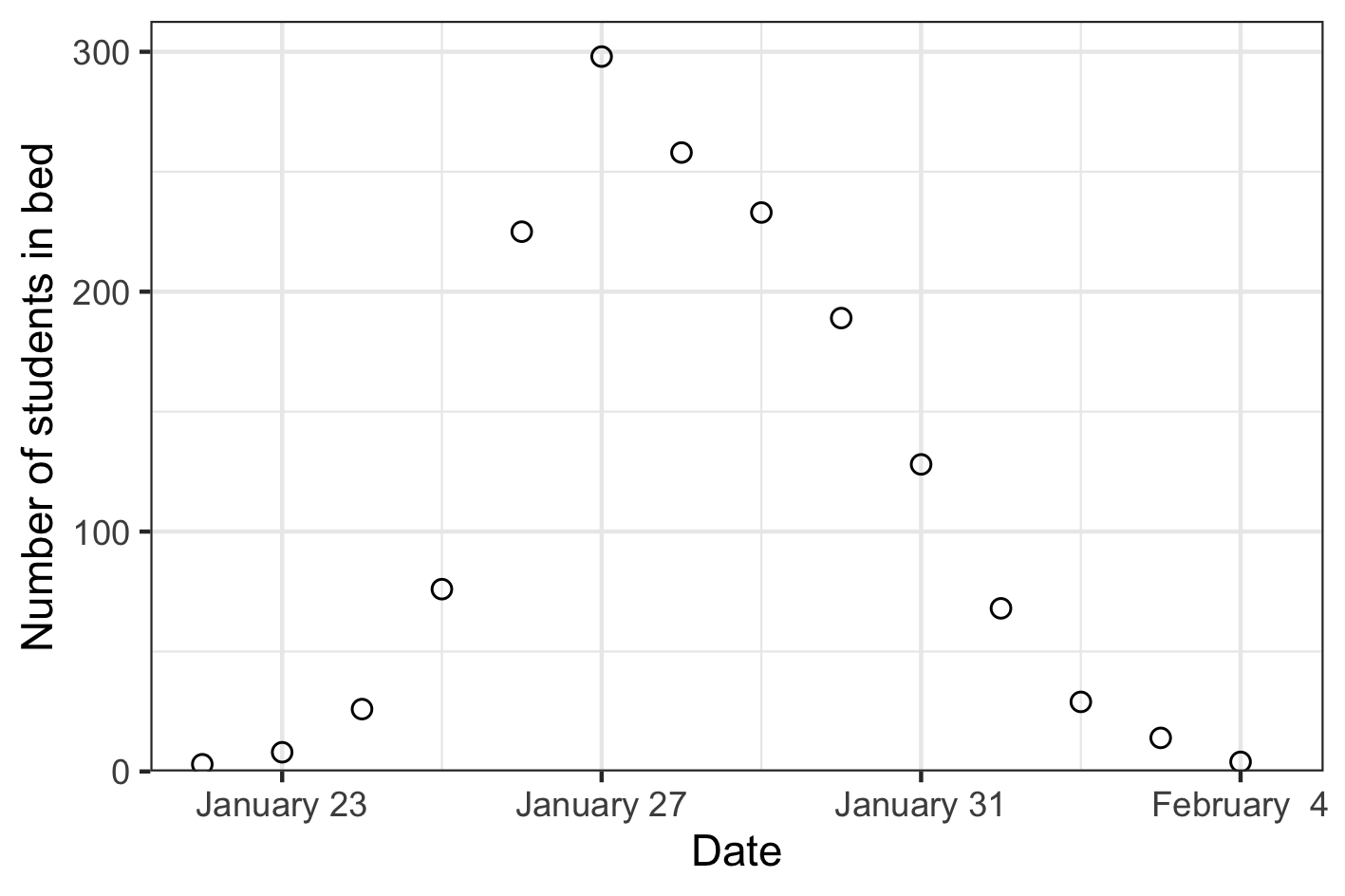}
    \caption{Number of students in bed each day during an influenza A (H1N1) outbreak at a British boarding school between January 22 and February 4, 1978.}
    \label{fig:influenza_data}
\end{figure}

In this section, we demonstrate how to use \texttt{Stan} and the Bayesian workflow on a simple example of disease transmission: an outbreak of influenza A (H1N1) in 1978 at a British boarding school.
The data consists of the daily number of students in bed, spanning over a time interval of 14 days, and is displayed in Figure \ref{fig:influenza_data}. There were 763 male students who were mostly full boarders and 512 of them became ill. The outbreak lasted from the 22nd of January to the 4th of February. One infected boy started the epidemic, which spread rapidly in the relatively closed community of the boarding school. 
The data are freely available in the \texttt{R} package \texttt{outbreaks}\cite{repidemics2020}, maintained as part of the R Epidemics Consortium.
The code from this analysis, including the \texttt{Stan} model, is available in the notebook online\footnote{\url{https://github.com/charlesm93/disease_transmission_workflow}}, which can serve as a hands-on tutorial to fit your own disease transmission model in \texttt{Stan}.

\subsection{Mathematical transmission model}
\begin{figure}
    \centering
    \includegraphics[scale=0.9]{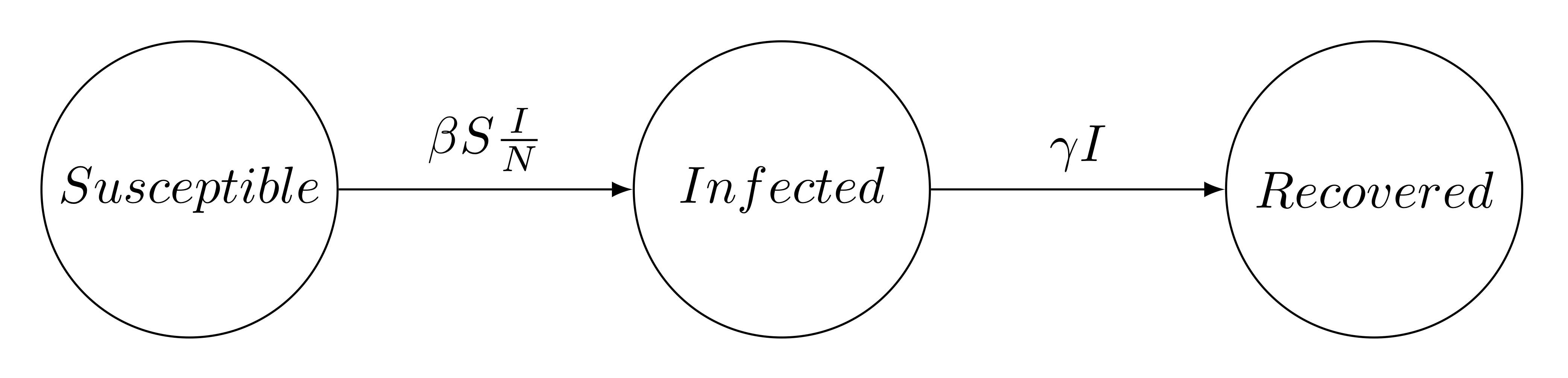}
    \caption{Diagram of the classic Susceptible-Infectious-Recovered (SIR) compartmental model. }
    \label{fig:sir_diagram}
\end{figure}

As discussed in section \ref{sec:intro}, epidemiological transmission models come in various forms, the two main categories being \textit{agent-based models}, which model each individual but are computationally expensive, and \textit{population-based models}, which stratify the population into several homogeneous compartments. This section focuses on one of the simplest population-based models, which is popular for its conceptual simplicity and its computational tractability: the Susceptible-Infected-Recovered (SIR) model. The SIR model splits the population into three time-dependent compartments: the susceptible, the infected (and infectious), and the recovered (and not infectious) compartments. When a susceptible individual comes into contact with an infectious individual, the former can become infected for some time, and then recovers and becomes immune.
The dynamic is summarized graphically in Figure \ref{fig:sir_diagram}.
The temporal dynamics of the sizes of each of the compartments are governed by the following  system of ODEs:
\begin{empheq}[left=\empheqlbrace]{align*} 
 \frac{dS}{dt} &= -\beta  S \frac{I}{N} \nonumber \\ 
 \frac{dI}{dt} &= \beta  S  \frac{I}{N} - \gamma  I \\ 
 \frac{dR}{dt} &= \gamma I \nonumber
\end{empheq}
where $S(t)$ is the number of people susceptible to becoming infected (no immunity), $I(t)$ is the number of people currently infected (and infectious), $R(t)$ is the number of recovered people (we assume they remain immune indefinitely), $\beta$ is the constant rate of infectious contact between people (often called transmission rate) and $\gamma$ is the rate of recovery of infected individuals (often called recovery rate, but formally relates to the duration of infectiousness rather than to the duration of symptoms).

The proportion of infected people among the population is $I / N$. At each time step, given uniform contacts, the probability for a susceptible person to become infected is thus $\beta I / N$, with $\beta$ the average number of contacts per person per time step multiplied by the probability of disease transmission when a susceptible and an infected subject come in contact. Hence, at each time step, $\beta  S I/N$ susceptible individuals become infected,  meaning $\beta  S I/N$ people leave the $S$ compartment and enter the $I$ compartment. Similarly, the recovery of an infected individual is taking place at rate $\gamma$, and thus the number of infected individuals decreases with speed $\gamma I$ while the number of recovered grows at the same speed.

The above model holds under several assumptions, making it most adapted to large epidemics occurring over relatively short periods of time: 

\begin{itemize}[noitemsep]
    \item births and deaths are not contributing to the dynamics and the total population $N=S+I+R$ remains constant, 
    \item recovered individuals do not become susceptible again over time,
    \item there is no incubation period before becoming infectious,
    \item the time spent in the infectious compartment follows an exponential distribution with mean $1 / \gamma$,
    \item the transmission rate $\beta$ and recovery rate $\gamma$ are constant in time (no behavioural changes nor control measures), 
    \item individuals may meet any other individual uniformly at random (homogeneous mixing),
    \item replacing the integer number of people in each compartment by a continuous approximation is legitimate since the population is big enough.
\end{itemize}

\subsection{Probabilistic transmission model}
We have described a deterministic transmission model, but we want to incorporate randomness into it to model our imperfect knowledge of the model parameters, natural variation and the imperfect fit of the model to reality. 
As discussed in section \ref{sec:bayes}, we need to specify a sampling distribution $p(\mathcal Y \mid \theta)$, and a prior distribution $p(\theta)$, with $\theta = \{\beta, \gamma, 1/\phi\}$ denoting all the parameters of the SIR model plus an additional over-dispersion parameter $\phi$ that is reparametrized as its inverse (see below). Given specific values for the parameters and initial conditions, 
a compartmental model defines a unique solution for each of the compartments, 
including the number of infected students at time $t$, $I(t)$. 
We want to link this solution to the observed data, i.e the number of students in bed  $\mathcal Y$.
We choose to model the number of students in bed with a count distribution that provides some flexibility regarding dispersion, the negative binomial distribution. 
This distribution allows us to use $I(t)$ as the expected value
and account for over-dispersion through parameter $\phi$:
$$
p(\mathcal Y \mid \theta) = \text{Negative-binomial}(\mathcal Y \mid I(t), \phi).
$$

We still need to specify a prior distribution for each of the three parameters using basic domain knowledge.
For the transmission rate we select 
$p(\beta) = \text{Normal}^+(2, 1)$
a \textit{weakly-informative prior} that only restricts $\beta$ to be positive (the half-normal distribution is truncated at 0) and puts a soft higher limit around 4 -- $p(\beta<4) = 0.975$.
For the recovery rate, we specify 
$p(\gamma) = \text{Normal}^+(0.4, 0.5)$, which expresses our belief that $\gamma$ has to be positive and that $p(\gamma \leq 1) = 0.9$ 
(i.e the probability that the average time spent in bed is less than 1 day is 0.9 \textit{a priori}). 
For the dispersion parameter, we use 
$p(1/\phi) = \text{exponential}(5)$ 
, as it is recommended in this situation to reparametrize the dispersion parameter as its inverse to avoid putting too much of the prior mass on models with a large amount of over-dispersion \cite{stan2020priorchoice}.
We can change these priors if more information becomes available, constraining our parameter estimation more tightly or, on the contrary, increasing its variance. See \cite{stan2020priorchoice} for more recommendations on prior choice. 

\subsection{Building an ODE-based model in \texttt{Stan}}

Writing this model in \texttt{Stan} is very similar to writing it mathematically.
As for any \texttt{Stan} model, we follow the block structure described in figure \ref{fig:blocks}.
In the \texttt{parameters} block, where we declare the model parameters \texttt{beta}, \texttt{gamma} and \texttt{phi\_inv}, using an inverse reparametrization for $\phi$ as discussed above, and their range of support -- strictly positive in these cases.
The full code can be found in the complementary notebook.
Here, we focus on solving the ODEs, which is the more intricate component of this model.
\texttt{Stan} supports numerical solvers for ODEs of the form $y' = f(t, y, \vartheta,x)$, where $t$ is usually time, $y$ the solution (i.e. the size of each compartment $\{S(t), I(t), R(t)\}$ in our example), and $y'$ the derivative of $y$ with respect to $t$.
Finally $\vartheta$ and $x$ encode additional inputs when evaluating $f$.

We specify $f$ inside the \texttt{functions} block, as follows:

\begin{lstlisting}[style=stan, numbers=none]
real[] f(real t, real[] y, real[] vartheta, real[] x_r, int[] x_i) {
	real S = y[1];
	real I = y[2];
	real R = y[3];
	real beta = vartheta[1];
	real <@\textcolor{black}{gamma}@> = vartheta[2];
	real N = x_i[1];
	real dS_dt = -beta * I * S / N;
	real dI_dt =  beta * I * S / N - <@\textcolor{black}{gamma}@> * I;
	real dR_dt =  <@\textcolor{black}{gamma}@> * I;
  	return {dS_dt, dI_dt, dR_dt};
}
\end{lstlisting}
The function $f$ returns $y'$ as an array of real numbers, as noted by \texttt{real[]}.
The first two arguments are the time, $t$, and the solution state, $y$.
The next three arguments are additional inputs.
These are concatenated inside arrays of real or integer numbers.
It is crucial to distinguish variables which are model parameters from variables which stay fixed as the sampler explores the parameter space.
\texttt{vartheta} contains variables that depend on the model parameters $\theta$, specifically $\beta$ and $\gamma$.
Quantities that influence the ODE but remain fixed during the whole procedure (e.g. the population size) are passed using \texttt{x\_r} and \texttt{x\_i}, depending on whether they are real or integers.
Distinguishing parameters and fixed variables is a matter of computational efficiency.
Section~\ref{sec:scaleup} discusses in details the computational cost of solving ODEs in \texttt{Stan}. 

Having defined the ODE system, we can obtain its approximate solution at any time point of interest by using a numerical solver.
In addition to parameters \texttt{theta} and constants \texttt{x\_r} and \texttt{x\_i}, the solver requires the initial time \texttt{t\_0}, the initial conditions \texttt{y\_0} and the different time points \texttt{ts} at which a solution is needed.
For this example, we use a Runge-Kutta integrator.
The call to the numerical solver is:
\begin{lstlisting}[style=stan, numbers=none]
  real[,] y = integrate_ode_rk45 (function f, real[] y0, real t0, real[] ts, 
                                  real[] theta, real[] x_r, int[] x_i);
\end{lstlisting}
The integrator returns a two-dimensional array with the solution of each compartment at each time point in \texttt{ts}.
Note that we pass to the integrator the arguments \texttt{theta}, \texttt{x\_r}, and \texttt{x\_i}, which are in turn passed to the function \texttt{f} each time the integrator is being called. As of \texttt{Stan} version 2.25, we can relax the signature of \texttt{f} and replace \texttt{theta}, \texttt{x\_r}, and \texttt{x\_i} with any convenient combination of arguments \cite{bales2020upgrading}.

\subsection{Criticizing the model before looking at the data}

\begin{figure}
	\centering
	\includegraphics[width=.9\linewidth]{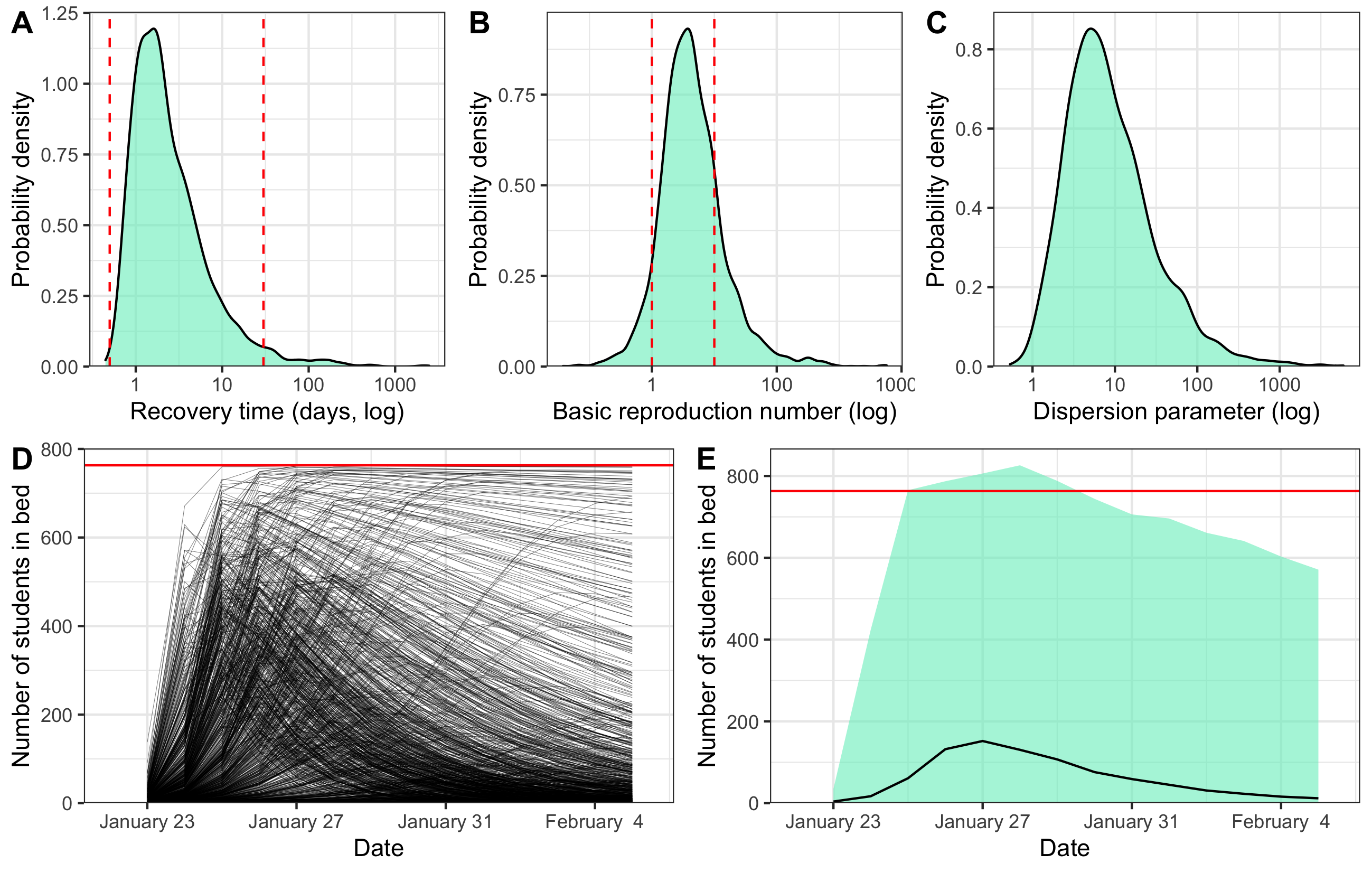}
	\caption{Prior predictive checks for (A) the recovery time ($1/\gamma$), (B) the basic reproduction number $\mathcal R_0$ ($\beta/\gamma$), (C) the dispersion parameter ($\phi$), (D) a set of 1,000 epidemic trajectories (each line is a unique simulated trajectory) and (E) the range in the numbers of students in bed each day (the line is the median and the light teal area is the 95\% central range). All these quantities are sampled from the prior distributions of the parameters. 		
	 The dashed red lines correspond to weak bounds from our domain knowledge where available:  the recovery time is expected to last between 0.5 and 30 days, $\mathcal R_0$ cannot be lower than 1 or higher than 10. The plain horizontal red line shows the population size (763).}
	\label{fig:priorpred}
\end{figure}

Before fitting the model to the data, it is useful to check that what we have encoded into the model is indeed coherent with our expectations. 
This is especially useful for complex models with many transformations. 
We can check if our priors are sound and correspond to domain knowledge by computing the \textit{a priori} probability of various epidemiological parameters of interest.
For instance for influenza, it is generally accepted that the basic reproduction number $\mathcal R_0$ is typically between 1 and 2, and that the recovery time is approximately 1 week. 
We want priors that allow for every reasonable configuration of the data 
but exclude patently absurd scenarios, per our domain expertise. 
To check if our priors fulfill this role, we can do a \textit{prior predictive check} \cite{gabry2019visualization}. More precisely, for any quantity of interest $\mathcal Y$, we can sample $\mathcal Y_\mathrm{prior}$ from the \textit{a priori} distribution of $\theta$, using the following sequential procedure:
\begin{eqnarray*}
  \theta_\mathrm{prior} & \sim & p(\theta), \\
  \mathcal Y_\mathrm{prior} & \sim & p(\mathcal Y \mid \theta_\mathrm{prior}).
\end{eqnarray*}
As demonstrated in the notebook\cite{grinsztajn2020bayesian}, doing so in \texttt{Stan} is very simple, as one can use almost the same model implementation that is used for inference. 
Figure \ref{fig:priorpred}A shows the distribution of the log of the recovery time, 
with the red lines showing loose bounds on the recovery time (0.5 and 30 days). 
We observe that most of the probability mass is between the red bars
but we still allow more extreme values, meaning our posterior can concentrate outside the bars, if the data warrants it. Figure \ref{fig:priorpred}B shows the same thing for $\mathcal R_0$, the loose bounds being 1 and 10 (note that for the SIR model, $\mathcal R_0 = \beta / \gamma$), and figure \ref{fig:priorpred}C for the dispersion parameter $\phi$.


Another aspect of prior predictive checking is to simulate fake data.
Using a similar procedure, we can simulate a set of ODE outputs compatible with the chosen prior distributions of parameters (Figure \ref{fig:priorpred}D), or, by adding the variability from the negative binomial distribution, the predicted range of the number of students in bed each day (Figure \ref{fig:priorpred}E).
It appears that the simulated trajectories are of the same magnitude as the data but still quite diverse, with some epidemics spreading quickly to almost every student and keeping most in bed for two weeks, and on the opposite several delayed or short-lived epidemics.
Looking at predicted range of students in bed, we observe that in some extreme situations this number can exceed the total number of students at the boarding school of 763.
This is caused by the fact that, if the SIR takes into account the population size, the negative binomial distribution that we put on top of it does not.
The prior predictive check often allows to detect this type of non-immediately obvious behaviors from the model.
In this case it should not have any consequence.

Put together, the prior predictive checks show that the model is not excessively constrained by the priors, and is still able to fit a wide variety of situations and data.
In fact, these priors are likely too wide and encompass scenarios that are impossible or extremely unlikely.
Typically, we can get away with priors that do not capture all our \textit{a priori} knowledge, provided the data is informative enough.
However when dealing with complicated models and relatively sparse data, 
we usually need well-constructed priors to regularize our estimates and avoid non-identifiability.
We conclude from the prior predictive checks that our choice of prior distributions is adequate.


\subsection{Fitting the model}
Once the model is written in \texttt{Stan} code, the fitting procedure is straightforward from \texttt{R} using the package \texttt{rstan}, or from one of the other available interfaces.
Using the default settings, we run 4 independent Markov chains, initialized at different random points, with a warm-up phase that lasts 1,000 iterations and a sampling phase that runs for another 1,000 iterations.
There is no definitive rule about the number of chains and iterations. 
The rule of thumb is to use the default settings as a starting point, and adapt these if needed after criticizing the inference.
An important point with regards to computing time is that the chains can be run in parallel, so it can make sense to run as many chains as there are processor cores and reduce the number of iterations per chain accordingly (down to a certain point).
For complex models it can be necessary to resort to high-performance computing clusters.

\subsection{Criticizing the inference}
After fitting the model, we check whether the inference is reliable.
In this setting, this means validating that the samples are unbiased and that our posterior estimates, based on these samples, are accurate.

\subsubsection{\texttt{Stan}'s diagnostics}
\texttt{Stan} provides a host of information to evaluate whether the inference is reliable. 
During sampling, \texttt{Stan} issues warnings when it encounters a numerical error: for example, failure to solve the ODEs or to evaluate the log joint distribution, $\log p(\mathcal Y, \theta)$.
A large number of such failures is indicative of numerical instability, which can lead to various kinds of problems: slow computation or inability to generate samples in certain regions of the parameter space.
These are often caused by coding mistakes, and the first step towards a solution is debugging the \texttt{Stan} code.
Fortunately, with the example at hand, we observe no such warning messages.

After the sampling is finished, \texttt{Stan} will run a set of basic diagnostics and issue a warning if it finds divergent transitions, saturated maximum tree depth or low energy\footnote{These diagnostics can also be accessed in \texttt{rstan} with the \texttt{check\_hmc\_diagnostics()} function.}.
Of these, the most serious is the presence of divergent transitions as it really challenges the validity of the inference, and requires revisiting the model or tweaking the inference settings.
Exceeding maximum tree depth is a concern about sampling efficiency rather than validity.
A low energy warning can be more problematic, as it suggests that the Markov chains did not explore the posterior distribution efficiently. 
Running the inference again with more iterations is often sufficient to solve this issue.
More details about \texttt{Stan}'s warnings and solutions are available in \cite{stan2020warnings}.

The summary table provides several quantities to troubleshoot the inference:
\begin{center}
\begin{lstlisting}[style=stan, numbers=none]
	         mean se_mean   sd 2.5%  25%  50%  75% 97.5% n_eff Rhat
	beta     1.73       0 0.05 1.63 1.70 1.73 1.77  1.84  2544    1
	<@\textcolor{black}{gamma}@>     0.54       0 0.05 0.45 0.51 0.54 0.57  0.63  2275    1
	phi_inv  0.14       0 0.07 0.04 0.08 0.12 0.17  0.33  2271    1
\end{lstlisting}
\end{center}
The $\widehat R$ statistics estimates the ratio between the overall variance and the within chain variance: if $\widehat R \not\approx 1$, then the Markov chains are not mixing and at least one of the chains is producing biased samples.
To measure how ``informative'' the samples are, we may use the \textit{effective sample size}, $n_\mathrm{eff}$, which is typically smaller than the total number of sampling iterations.
Samples from an MCMC procedure tend to be correlated, meaning there is redundant information in the samples which makes the posterior estimates less precise.
Conceptually, the variance of our Monte Carlo estimates corresponds to the variance we would expect if we used $n_\mathrm{eff}$ independent samples.
Stan's procedure for computing $\widehat R$ and $n_\mathrm{eff}$ is based on recent improvements made to both estimates\cite{vhetari2020rhat}.

\begin{figure}
	\centering
	\includegraphics[width=.9\linewidth]{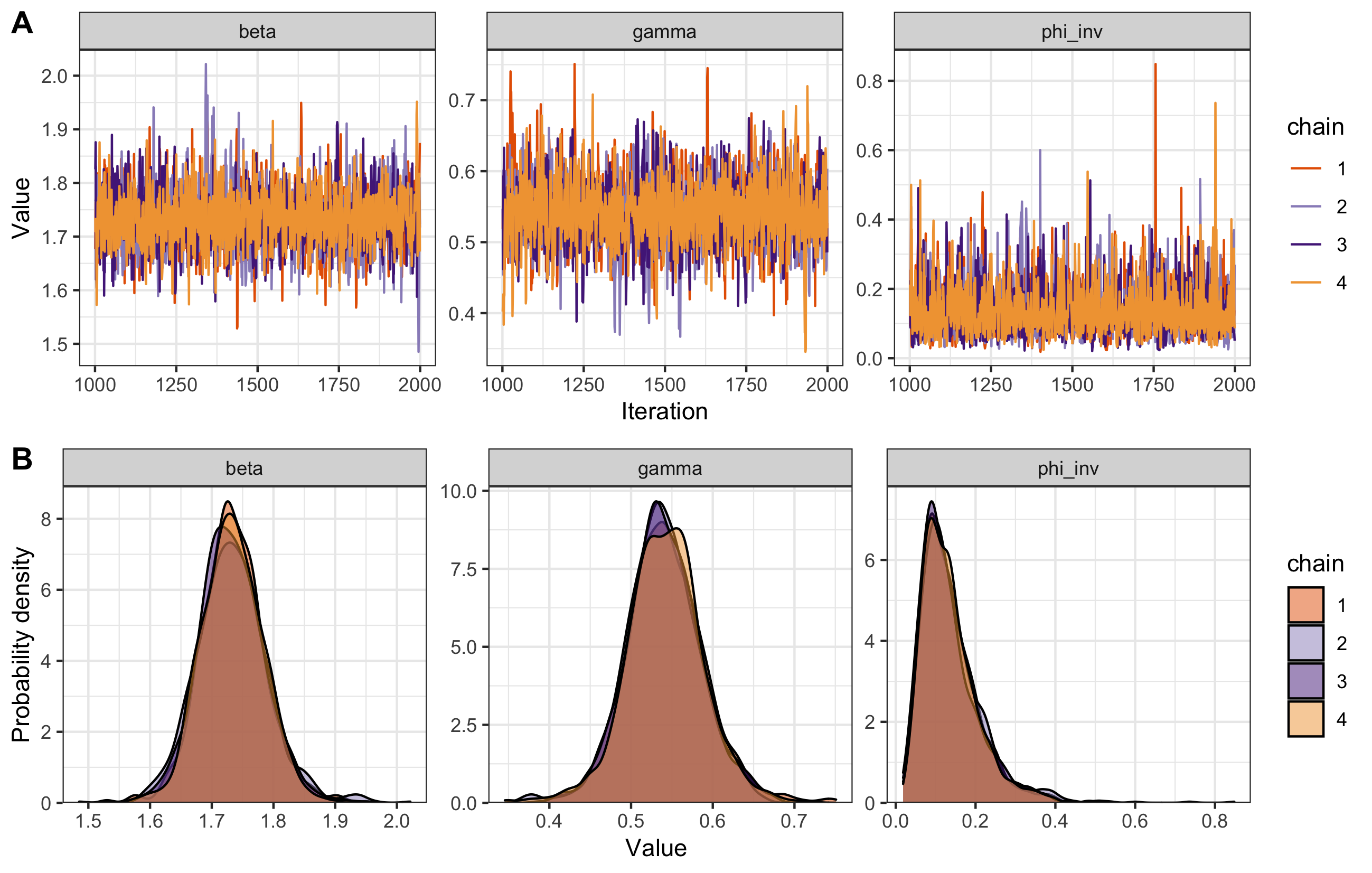}
	\caption{(A) Trace plot showing the value of each chain at each iteration (excluding warm-up) and (B) marginal posterior densities for the transmission rate (\texttt{beta} or $\beta$), the recovery rate (\texttt{gamma} or $\gamma$) and the inverse dispersion parameter (\texttt{phi\_inv} or $1/\phi$), separately for each of the four Markov chains.}
	\label{fig:chain_densities}
\end{figure}

Here we note that $\widehat R$ is close to 1 ($<$ 1.01) and that \texttt{n\_eff} is large, which makes us confident we can rely on the inference.
Conversely, large $\widehat R$ and low \texttt{n\_eff} would indicate that the Markov chains are not cohesively exploring the parameter space and, in turn, that our estimates of the posterior mean and quantiles are unreliable.
Having no \texttt{Stan} warning is a good, if imperfect, indication that the inference is reliable. 
We can furthermore plot the ``trace'' (Figure \ref{fig:chain_densities}A) and the marginal posterior densities (Figure \ref{fig:chain_densities}B) of each chain to confirm that the Markov chains mix well and are in agreement with one another.

When the diagnostics reveal a problem with our inference, we must consider several sources of error.
A common, if trivial, problem is coding errors.
If we are confident in our implementation, we may inspect our model specification.
Often times, reparameterizing the model\cite{stan2020reparametrization} or using stronger priors, when the requisite information is available, improves the interaction between HMC and the model, leading to better mixing of the chains and moreover better inference.
Sometimes the model at hand, i.e. the likelihood and the prior, is the model we need to fit and we must entertain the possibility that our inference engine is not suited for our problem.
In this case, we may consider changing the tuning parameters of HMC or all together adopt a new inference scheme,
e.g. marginalization\cite{rue2009inla, rue2017inla, margossian2020laplace}, variational inference\cite{blei2017vi}, or adaptive Metropolis and Gibbs\cite{roberts2009adaptive}, to only name a few popular approaches.
Indeed HMC is not a one-size-fits-all solution and, depending on the context, other techniques can offer better performance.
The principles of the Bayesian workflow apply to other inference strategies, although we should be mindful that not all algorithms are amiable to diagnostics and that some techniques can require extensive revision whenever we change model (or even be restricted to a narrow menu of models).

%
%

\subsubsection{Criticize the inference with simulated data}
While there exist many theoretical guarantees for MCMC algorithms,
modelers should realize that these rely on a set of assumptions
which are not always easy to verify
and that many of these guarantees are \textit{asymptotic}.
This means they are proven in the limit where we have an infinite number of samples from the posterior distribution.
A very nice, if advanced, review on the subject can be found in \cite{Roberts2004mcmc}.
As practitioners, we must contend with finite computational resources
and assumptions which may or may not hold.
The diagnostics we reviewed earlier -- e.g. $\widehat R$ or effective sample sizes --
provide \textit{necessary} conditions for
the MCMC sampler to work but not \textit{sufficient} ones.
Nevertheless, they are potent tools for diagnosing shortcomings in our inference.
This subsection provides further such tools,
from both a rigorous and a pragmatic perspective.

Fitting the model to simulated data is, if done properly, an effective way
to test whether our inference algorithm is reliable.
If we cannot successfully fit the model in a controlled setting,
it is unlikely we can do so with real data.
This of course raises the question of what is meant by ``successfully fitting''
the model.
In a Bayesian setting, this means our inference procedure accurately estimates various characteristics of the posterior distribution, such as the posterior mean, variance, covariance, and quantiles.

A powerful method to check the accuracy of our Bayesian inference
is \textit{simulation-based calibration} (SBC) \cite{talts2020sbc}.
SBC exploits a very nice consistency result.
The intuition is the following: if we draw several sets of parameters from our prior distribution
$$
  \theta_1, ..., \theta_2 \sim p(\theta)
$$
and for each set of parameters $\theta_i$ simulate a data set $\mathcal Y^i$, we can by fitting the model multiple times recover the prior distribution
from the estimated posteriors. This technique is a bit beyond the scope of this tutorial,
though we vividly encourage the reader to consult the original paper.

For the time being, we focus on a simpler heuristic: fit the model to one simulated data set and check if we recover the correct parameter values.
There are serious limitations with this approach: when do we consider that the estimated posterior distribution covers the correct value or how do we know if the variance of the posterior is properly estimated? 
But the test is useful: in this controlled setting, do the chains converge? Is the computation of the log density numerically stable (e.g. are we using the right ODE integrator)? Do our priors prevent the chains from wandering into absurd regions
of the parameter space? These are all questions this simple test can help us tackle.

\begin{figure}
	\centering
	\includegraphics[width=.8\linewidth]{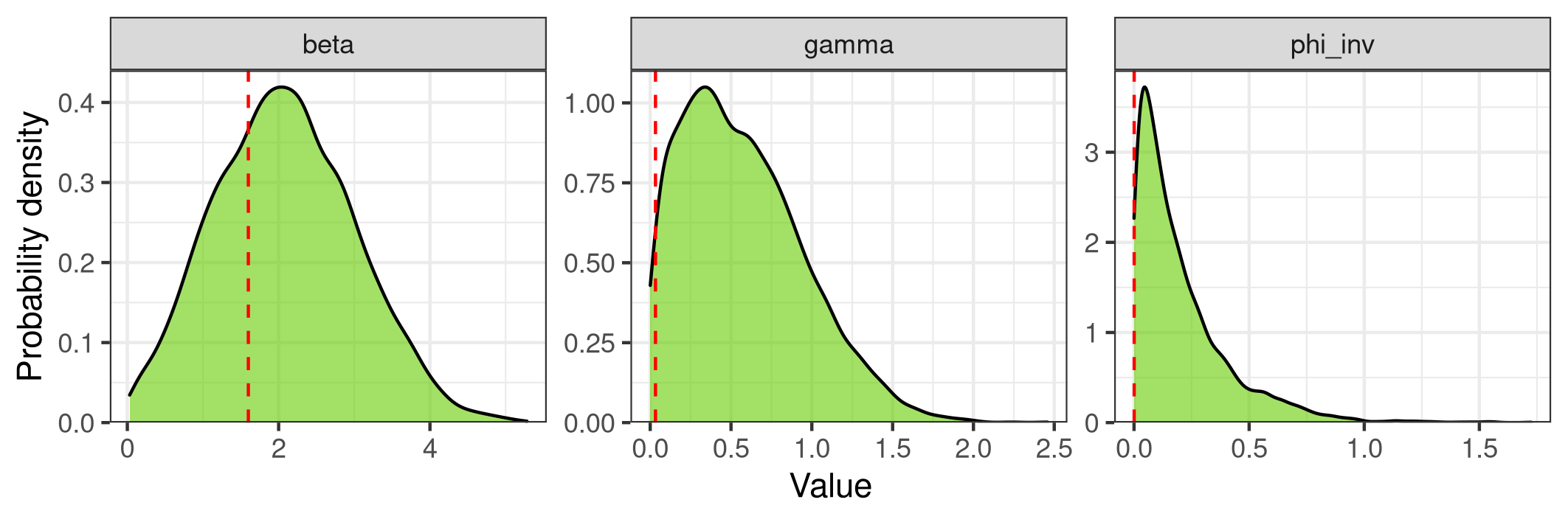}
	\caption{Marginal posterior densities for the transmission rate (\texttt{beta} or $\beta$), the recovery rate (\texttt{gamma} or $\gamma$) and the inverse dispersion parameter (\texttt{phi\_inv} or $1/\phi$) obtained when fitting the model to simulated data. The red dashed lines show the fixed parameter values used for simulating the data.}
	\label{fig:inference_simu}
\end{figure}

We take one arbitrary draw from the prior distribution and use it as data to which we fit the model.
In this specific case, we used the following randomly-selected values: $\beta = 1.6$, $\gamma=0.033$ and $1/\phi=0.0007$.
We then plot the estimated posterior distribution (Figure \ref{fig:inference_simu}) along with the ``true'' parameter values.
The density function covers the true parameters, although it is not always centered on it.
The latter is not alarming, especially if the parameter values that were selected
lie on the tail of the prior distribution. We could repeat this process a few times to get a better sense of the performance of the model together with the inference algorithm.

\subsection{Criticize the fitted model}


Now that we trust our inference, we can check the utility of our model.
Utility is problem-specific and can include the
precise estimation of a parameter or predicting future behaviors.
In general, it is good to check if our model, once fitted, produces simulations
that are consistent with the observed data.
This is the idea behind \textit{posterior predictive checks}.

\begin{figure}[h]
	\centering
	\includegraphics[width=.8\linewidth]{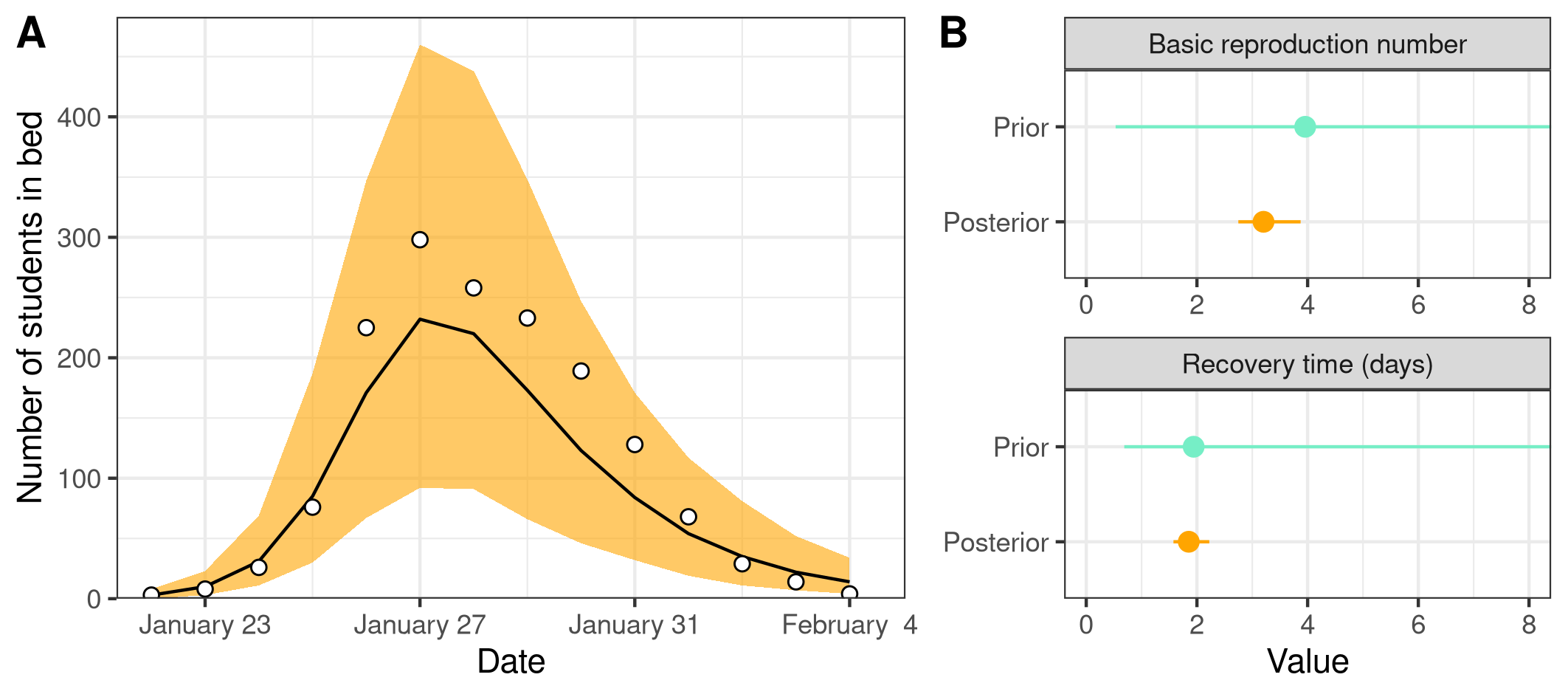}
	\caption{(A) Posterior predictive check of the number of students in bed each day during an influenza A (H1N1) outbreak at a British boarding school. The line shows the median and the orange area the 90\% prediction interval. (B) Prior and posterior predictive checks of the basic reproduction number $\mathcal R_0$ and of the recovery time (both truncated at 8). The dot shows the median posterior and the line shows the 95\% credible interval.}
	\label{fig:ppc_students}
\end{figure}

We sample predictions, $\mathcal Y_\mathrm{pred}$, using the following sequential procedure:
\begin{eqnarray*}
  \theta_\mathrm{post} & \sim & p(\theta \mid \mathcal Y), \\
  \mathcal Y_\mathrm{pred} & \sim & p(\mathcal Y \mid \theta_\mathrm{post}).
\end{eqnarray*}
This is identical to prior predictive checks, except that we are now sampling parameters from the posterior distribution, rather than from the prior.
We use these samples to construct a fitted curve for students in bed,
together with a measure of uncertainty (e.g. the 95\% prediction interval, 
meaning that observed data points are expected to fall outside of the interval once every twenty times).
The posterior predictive check represented in Figure \ref{fig:ppc_students}A allows us to confirm that the model fits the data reasonably well.
We can also do separate posterior predictive checks for each Markov chain.

At this point, we can confidently move forward with the utility of the model, for instance conclude that the basic reproduction number of this outbreak can be estimated to 3.2 with a 95\% credible interval of 2.7 - 3.9 (respectively from the median, 2.5\% quantile and 97.5\% quantile of the posterior samples), or that the average recovery time can be estimated to 1.8 days with a 95\% credible interval of 1.6 - 2.2 (Figure \ref{fig:ppc_students}B).

%% file: scaleup.tex
 
Doing MCMC on ODE-based models can be computationally intensive, especially as we scale up the number of observations, parameters, and start using more sophisticated ODEs.
If we want to reap the benefits of a full Bayesian inference, we have to pay the computational cost.
But while we cannot get away with a free lunch, we can avoid an overpriced one.
\texttt{Stan} is a flexible language, which means there are worse and better ways of coding things.

This section develops a few principles to make ODE models in \texttt{Stan} more scalable, drawing on our experience with an SEIR model of SARS-CoV-2 transmission\cite{riou2020covid}.
The key is to understand the computational cost of each coding block (Section~\ref{sec:blocks}) and recognize that operations in the \texttt{transformed parameters} and \texttt{model} blocks dominate the computation.
A brief primer on the mechanism of HMC will make this clear.
Our goal is then to write our model in such a way that we limit the number of operations in the expensive blocks.
Another important consideration is the need to solve and differentiate the ODEs for our transmission model many times.
We present ways to make numerical integration cheaper.
Anecdotally, implementing the strategies we outline here reduced the runtime of our model from three days to two hours!
These techniques will also be much help in our next example in section~\ref{sec:extension}.

\subsection{The Computational cost of Stan's coding blocks}

\texttt{Stan} abstracts the inference away from the modeling but it's worth taking a peek inside the black box.
HMC simulates physical trajectories in the parameter space by solving Hamilton's equations of motion -- in this instance, we may think of these equations as a convenient reformulation of Newton's law of motion.
Unlike random walk algorithms (e.g. Metropolis-Hastings, Gibbs), HMC does not start each iteration with a random step, but rather with a random momentum, given to a fictitious particle.
The acceleration of the particle is then driven by the gradient of the log posterior.
We solve the equations of motion numerically, using a leapfrog integrator.
More details can be found in references\cite{hoffman2014nuts, betancourt2018conceptual}.

Recall the \texttt{model} block specifies the joint distribution $\log p(\mathcal Y, \theta)$.
We further require the gradient
%
$$
  \nabla_\theta \log p(\mathcal Y, \theta) = \nabla_\theta \log p(\theta \mid \mathcal Y).
$$
Fortunately, the user does not need to specify the gradient.
Instead, \textit{automatic differentiation} generates the requisite derivatives under the hood using a numerical integrator
\cite{margossian2019ad, griewank2008ad}.
At each step the leapfrog integrator takes -- that is, multiple times per iteration -- we need to \textit{evaluate and differentiate} $\log p(\mathcal Y, \theta)$.

This perspective informs how each model block scales (Figure \ref{fig:blocks}).
The \texttt{data} and \texttt{parameters} blocks are used to declare variables.
The \texttt{transformed data} block is evaluated once.
The \texttt{transformed parameters} and \texttt{model} blocks are evaluated and differentiated at each integration step, which is multiple times per iteration.
The \texttt{generated quantities} block is evaluated once per iteration.
Hence operations in the \texttt{parameters} and \texttt{transformed parameters} blocks dominate the computational cost and should only entail operations that depend on $\theta$ and are required to compute $\log p(\mathcal Y, \theta)$.

In the previous example, solving the ODEs at the observed times is required to compute the likelihood and occurs in the \texttt{transformed parameters} block.
On the other hand, the computation of $\mathcal R_0$ is relegated to \texttt{generated quantities}.
Even though we want samples from $p(\mathcal R_0 \mid \mathcal Y)$ and $\mathcal R_0$ depends on the model parameters,
$\mathcal R_0$ does not contribute to $\log p(\mathcal Y, \theta)$.


\subsection{Reducing the cost of propagating derivatives through the ODE solution}

To obtain the requisite gradient, we must propagate derivatives through the ODE solution.
This differentiation process can become very expensive: understanding its mechanism can help us avoid certain computational pitfalls.


Our ODE is defined by
$$
  \frac{\mathrm d y}{\mathrm d t} = f(y, t, \vartheta, x).
$$
Here, $\vartheta$ contains inputs to $f$ that depend on the model parameters, $\theta$, while $x$ contains inputs which do \textit{not} depend on $\theta$ and therefore remain fixed as the Markov chain moves through the parameter space.
Note that in general, $\vartheta \neq \theta$.
In the SIR model for example, $\vartheta = \{\beta, \gamma\}$, while $\theta = \{\beta, \gamma, 1/\phi\}$.
To define the integral, we additionally specify an initial time $t_0$, times of integration $t_s$, and an initial condition $y_0$, all of which can vary with $\theta$.
Hence when propagating derivatives to compute the gradient of the log joint density, we need to worry about how the solution varies with respect to $\vartheta$ and potentially $y_0$\footnote{
While $t_0$ and $t_s$ may depend on model parameters, we here assume they do not to avoid some minute technicalities and simplify our discussion.}.
We call these elements \textit{varying parameters} and denote $K$ the number of such elements.
Furthermore, let $N$ be the number of \textit{states}, that is the length of $y$ or the number of compartments in a SIR-type model.

In \texttt{Stan}, we propagate derivatives by solving a coupled system of ODEs.
The intuition is the following.
Suppose we want to compute
$$
  \frac{\mathrm d y}{\mathrm d \vartheta}.
$$
We do not have an analytical expression for $y$, so a direct application of automatic differentiation is not feasible.
But we can, assuming the requisite derivatives exist, compute
$$
  \frac{\mathrm d f}{\mathrm d \vartheta} = \frac{\mathrm d}{\mathrm dt} 
    \frac{\mathrm dy}{\mathrm d \vartheta}
$$
and then integrate this quantity numerically.
The end result is that, instead of only solving for the $N$ original states, we solve an $N + NK$ system to both evaluate and differentiate $y$.\footnote{
Strictly speaking, we do not need to explicitly compute  $\mathrm d y / \mathrm d \vartheta$ to propagate derivatives; this is an important, if somewhat counter-intuitive, result of automatic differentiation, and motivates a so-called \textit{adjoint method}, which only requires solving $2N + K$ ODE states, albeit incurring additional overhead cost.
For a deeper discussion on the topic, we recommend \cite[Hindmarsh and Serban, 2020][]{hindmarsh2020ode}.}
Evidently, solving and differentiating the ODE is much more expensive than only solving it!
We this in mind, our goal should be clear: make $N$ and $K$ as small as possible.

The number of states $N$ in SIR-type models is simply the number of compartments and not much can be done to reduce this number.
Limiting $K$ is first and foremost a matter of book-keeping.
Recall the function signature of \texttt{Stan}'s numerical integrator:
\begin{lstlisting}[style=stan, numbers=none]
real[,] y = integrate_ode_rk45 (function f, real[] y0, real t0, real[] ts, 
		real[] theta, real[] x_r, int[] x_i);
\end{lstlisting}
For every element in $\vartheta$, we add an additional $N$ states to solve for.
Hence, components which do not depend on $\theta$ should be passed through \texttt{x\_r} and \texttt{x\_i}\footnote{Stan now offers a variadic signature, which is more flexible and automatically recognizes if an argument is parameter dependent or not\cite{stan2021manual}.}.

Suppose our initial condition, $y_0$, are varying parameters, i.e. depend on the model parameters.
It is not uncommon for some of the elements in $y_0$ to not depend on $\theta$.
For example, in a compartment model, the initial condition for the $I$ compartment may depend on model parameters,
while it is set to 0 for the $R$ compartment.
More generally, $y_0$ may only depend on $k < N$ parameters.
The straightforward approach is to pass $y_0$ as a vector of parameters.
\texttt{Stan} interprets this as $N$ additional varying parameters, which means the number of ODE we solve increases by $N^2$.
This is overpriced lunch!

A better, if more intricate, approach is to solve the ODEs for deviations from the baseline and split $y_0$ between $\vartheta$ and \texttt{x\_r}.
Concretely, let 
$$ 
  z = y - y_0 .
$$
The initial condition for $z$ is then $\bf 0$, an $N$-vector of 0's and a fixed quantity.
Now,
$$
  \frac{\mathrm d z}{\mathrm d t} = \frac{\mathrm d y}{\mathrm dt} = f(z + y_0, t, \vartheta, x)
    = \widetilde{f}(z, t, \widetilde \vartheta, \widetilde x)
$$
where $\widetilde f$ is the same map as $f$, but applied to $z + y_0$ instead of $z$;
$\widetilde \vartheta$ contains $\vartheta$ and the elements of $y_0$ that depend on $\theta$;
and $\widetilde x$ contains $x$ and the elements of $y_0$ that are fixed.
With this implementation, $K$ is kept to a minimum.
We recover the original $y$ simply by adding $y_0$ to $z$.
In the SEIR model by \cite[Hauser et al, 2020][]{riou2020covid}, we have 58 initial conditions but together these depend on a single parameter.
$\vartheta$ itself contains 4 elements.
Reparameterizing the ODE means we go from $K = 62$ to $K = 5$, that is from solving 3596 coupled ODEs to only solving 290.

\subsection{Picking the right ODE integrator}

The task of solving and differentiating an ODE boils down to integrating an augmented ODE system.
The majority of the time, we deal with nonlinear ODEs with no analytical solution and must resort to numerical integrators.
Hence to ensure reasonable performance, it is crucial to pick the right integrator and tune it properly.

Conceptually, numerical integrators perform a linear approximation of the solution between time points $t$ and $t + \delta t$, using the tangent, $f(t)$.
The step size $\delta t$ controls the usual trade-off between accuracy and speed, with a smaller $\delta t$ leading to more accurate results but slower computation.
The step size is adaptively computed by the integrator in order to control the target approximation error.
When calling an integrator in \texttt{Stan}, the user may specify the target error via
\begin{enumerate}
  \item the \textit{absolute tolerance}, which sets the upper bound for the absolute error in a solution,
  \item the \textit{relative tolerance}, which sets the upper bound for the error relative to the solution.
\end{enumerate}
Lower tolerance typically induces smaller step sizes.
Unless the user has a strong grasp on how these tuning parameters may affect the inference, we recommend using \texttt{Stan}'s default.
Another important tuning parameter is the \textit{maximum number of steps}, after which the integrator gives up on solving the ODE.
In a Bayesian context, it can be useful to end the integrator early, especially when the Markov chain wanders into odd regions of the parameter space, that make the ODE difficult to solve.
Such a wandering can occur during the warmup phase of HMC, before the chains converge to the relevant region of the parameter space.

\texttt{Stan} supports two ODE integrators\cite{stan2021manual}: a Runge-Kutta (RK45) method for non-stiff systems with
\texttt{
integrate\_ode\_rk45},
and a backward differentiation (BDF) algorithm for stiff systems with
\texttt{
integrate\_ode\_bdf}.
There is no formal definition of stiffness but the general idea is that the phenomenon occurs when the step size, $\delta t$, of the integrator needs to be extremely small -- smaller than what is needed to achieve the required accuracy -- in order to make the integrator stable.
Stiffness can arise when the scale of the solution varies largely as a function of $t$.
The RK45 integrator is typically faster, so we recommend it as a starting point.
If however the system is stiff, the RK45 integrator will be slow and numerically unstable and in this case, the BDF integrator is preferable.

Users should therefore be prepared to adjust their solver.
This can mean switching from RK45 to BDF, or adjusting the tuning parameters of the integrator.
When an integrator fails to solve an ODE, \texttt{Stan} issues a warning message and rejects the iteration being computed.
An excessive number of such failures can indicate the integrator needs to be adjusted.

For certain problems, knowing ahead of time if a system is stiff may not be obvious.
What is even less obvious is whether a coupled system is stiff.
And what is yet again less obvious is whether a system is stiff or nonstiff across the range of parameters the Markov chain explores, both during the warm-up and the sampling phases.
As said, the chain can, during the warmup phase, land in extreme regions of the parameter space and the ODE can become tremendously difficult to solve, leading to slow computation and numerical instability.
In our experience, this unfortunate behavior is often difficult to avoid.
Using more informative priors, when such information is available, can help the chains converge faster to ``sensible'' regions of the parameter space.
Carefully picking initial values for the Markov chain can also be helpful.


%% file: extension.tex
In section \ref{sec:simplesir}, we demonstrated how to use \texttt{Stan} to build a basic disease transmission model on the example of influenza data. We now show how to build a more complex model of SARS-CoV-2 transmission in Switzerland. 
We start with the SIR model from section \ref{sec:simplesir}, and then gradually build on it by iteratively increasing the complexity of the model and keeping in mind the principles of the Bayesian workflow described in section \ref{sec:workflow}. At each step, we aim to identify and eliminate a specific shortcoming. All obstacles can be roughly classified as inference issues or modeling issues. Inference issues can be, in most cases, diagnosed by \texttt{Stan}. They include errors in code, biases in MCMC, or unidentifiable parameters. Modeling issues include misspecified or unrealistic models that pass the inference diagnostics but do not describe the data well or lead to impossible parameter values. These can be detected by prior and posterior predictive checks. All through, we keep the \textit{folk’s theorem} \cite{gelman2020bayesian} in mind, according to which an inference problem is often due to a modeling problem.
For example, poorly specified priors can put probability mass in numerically unstable regions of the parameter space and frustrate our inference algorithms.
Table \ref{tab:table_covid} presents every iteration of our model.
The complete analysis, including the \texttt{Stan} code for the listed models, is available in a complementary notebook\cite{grinsztajn2020bayesian}. Below we only present a broad outline of the iterative modeling process.

\input{table_covid_extension.tex}

\subsection{Data}
\begin{figure}[h]
    \centering
    \includegraphics[width=.5\linewidth]{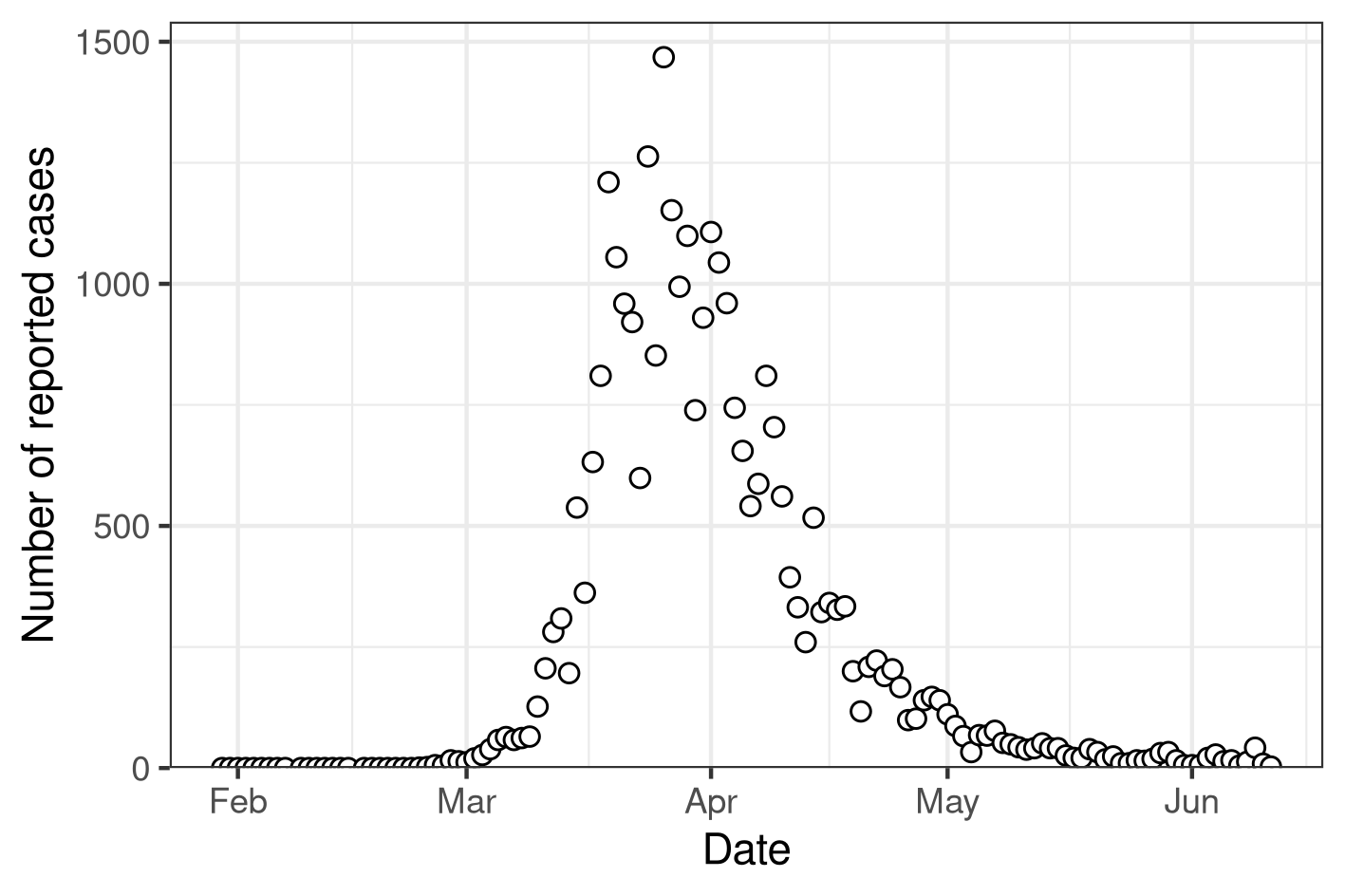}
    \caption{Daily number of reported cases of SARS-CoV-2 infection in Switzerland between February and June, 2020.}
    \label{fig:data_covid}
\end{figure}

The main dataset used in this section is the daily number of reported cases of SARS-CoV-2 infection in Switzerland at the national level during the first epidemic wave, from February to June 2020 (Figure \ref{fig:data_covid}). At a later modeling stage, we include serological antibody survey data from Geneva, available for dates between April 6th and May 9th 2020\cite{stringhini2020seroprevalence}.

%
%
%

\subsection{First attempt}
\label{sec:firstattempt}
As a first step, we directly try to fit the SIR model from section \ref{sec:simplesir} to this new data (model iteration \#1 in Table \ref{tab:table_covid}). We do not expect this simple model to perform well, but rather aim to create a baseline. We first need to account for one significant difference between influenza and SARS-CoV-2 datasets: the former represents prevalence (the number of students in bed being interpreted as a proxy for currently infected individuals), and the latter represents incidence (the number of new cases on a given day). To adjust for this discrepancy, we compute the incidence $\Delta I(t)$ from the compartments of the SIR model as the number of individuals entering the $I$ compartment during day $t$. The new sampling distribution is:
$$
p(\mathcal Y \mid \theta) = \text{Negative-binomial}(\mathcal Y \mid \Delta I(t), \phi)
$$
where $\mathcal Y$ now denotes data on reported cases.

\begin{figure}[h]
 	\centering
 	\includegraphics[width=.8\linewidth]{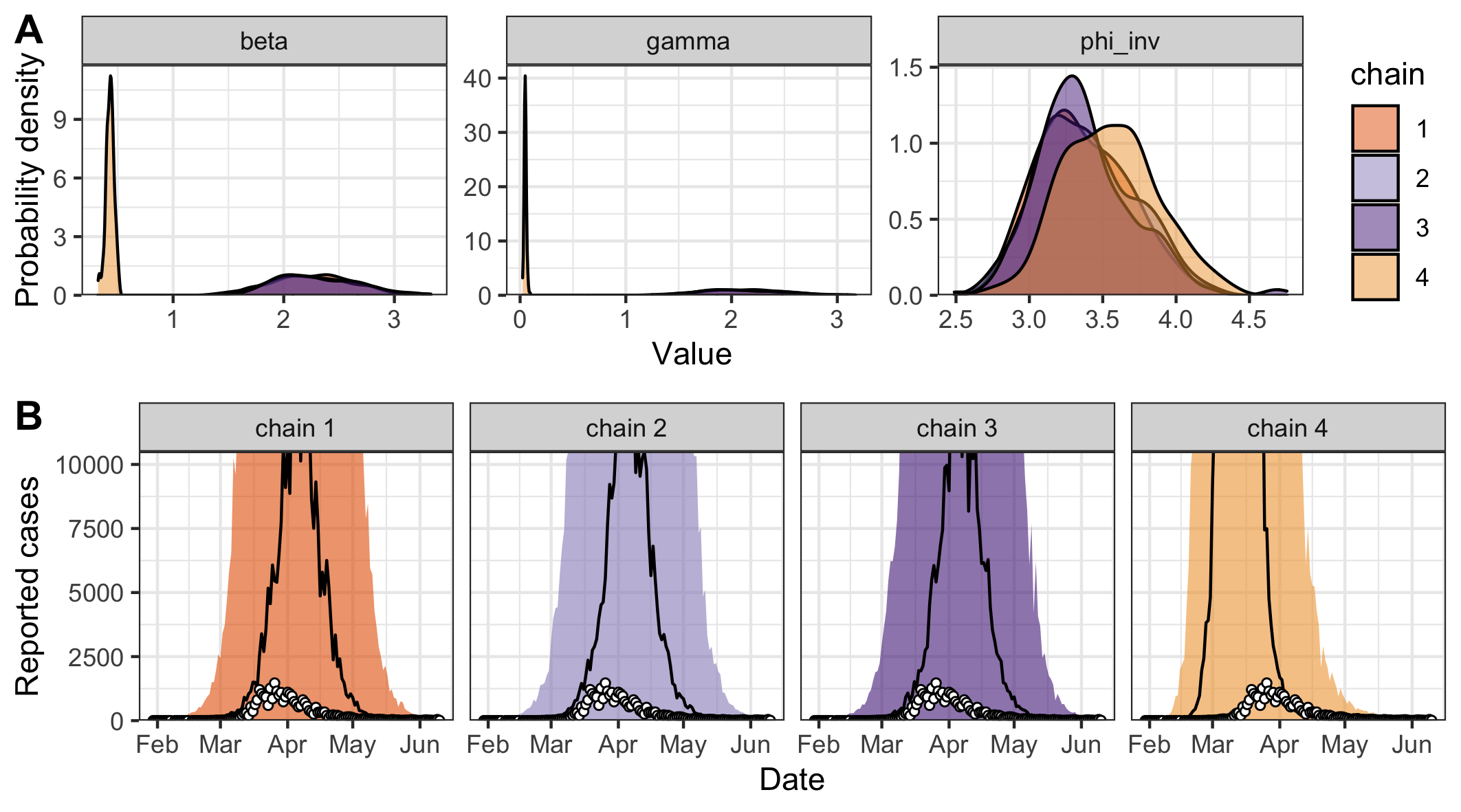}

    \caption{(A) Posterior distributions of the model parameters (the transmission rate $\beta$, the recovery rate $\gamma$ and the inverse dispersion parameter $1/\phi$) and (B) chain-by-chain posterior predictive check of the number of reported cases for the simple SIR model (model iteration \#1) applied to data on the SARS-CoV-2 epidemic in Switzerland (white circles).}
    \label{fig:ppc}
\end{figure}

Several diagnostics indicate that the inference from this first model fit should not be trusted: \texttt{Stan} issues a warning about divergent transitions, $\hat R$ is far larger than 1, and the effective sample size is very small.
We can also see that the Markov chains do not mix (Figure \ref{fig:ppc}A), and a posterior predictive check further shows that the model is, across all chains, unable to fit the data (Figure \ref{fig:ppc}B).
It is not uncommon for problems to come in bulk.
This is actually a feature of \texttt{Stan}: when it fails, it fails loudly.


\subsection{Model improvement}
Drawing from domain knowledge, we take several steps to improve the model.

\subsubsection{From SIR to SEIR with underreporting}
First, we add a reporting rate parameter, $p_r \in (0, 1)$, to account for the underreporting of cases (model iteration \#2). 
Indeed, contrary to the controlled environment of the boarding school where every student can be monitored, reported cases are an incomplete measurement of the true incidence of SARS-CoV-2. 
In order to be reported as cases, individuals infected with SARS-CoV-2 have to get tested but unfortunately not everyone gets tested; rather, individuals with severe symptoms or a perceived risk of severe outcome are more likely to be tested\cite{lipsitch2015potential}.
We furthermore need to account for the test sensitivity, i.e. is the test positive given the tested individual is infected?
All these considerations are captured by a single parameter, $p_r$, which we interpret as the probability that an infected individual is identified as infected, and for which we assign a weakly informative prior, $p(p_r) = \mathrm{Beta}(1, 2)$.
We still ignore the fact that reporting may take a few days, and assume that infected individuals appear as reported when they become infectious, i.e. as then enter compartment $I$.

Second, we add an \textit{Exposed} compartment $E$ to account for individuals who have been exposed to the virus but are not yet infectious. 
This leads to an SEIR model (model iteration \#3) that features another additional parameter: the \textit{incubation rate} $a$, defined as the inverse of the average incubation time.
We select a weakly-informative prior for this new parameter, $p(a) = \mathrm{Normal}^+(0.4, 0.5)$, which encodes the belief that exposed individuals become infectious after a period that lies between 0.5 and 30 days. 

\subsubsection{Limitations to the SEIR model with underreporting}
\begin{figure}[h]
		\centering
		\includegraphics[width=.8\linewidth]{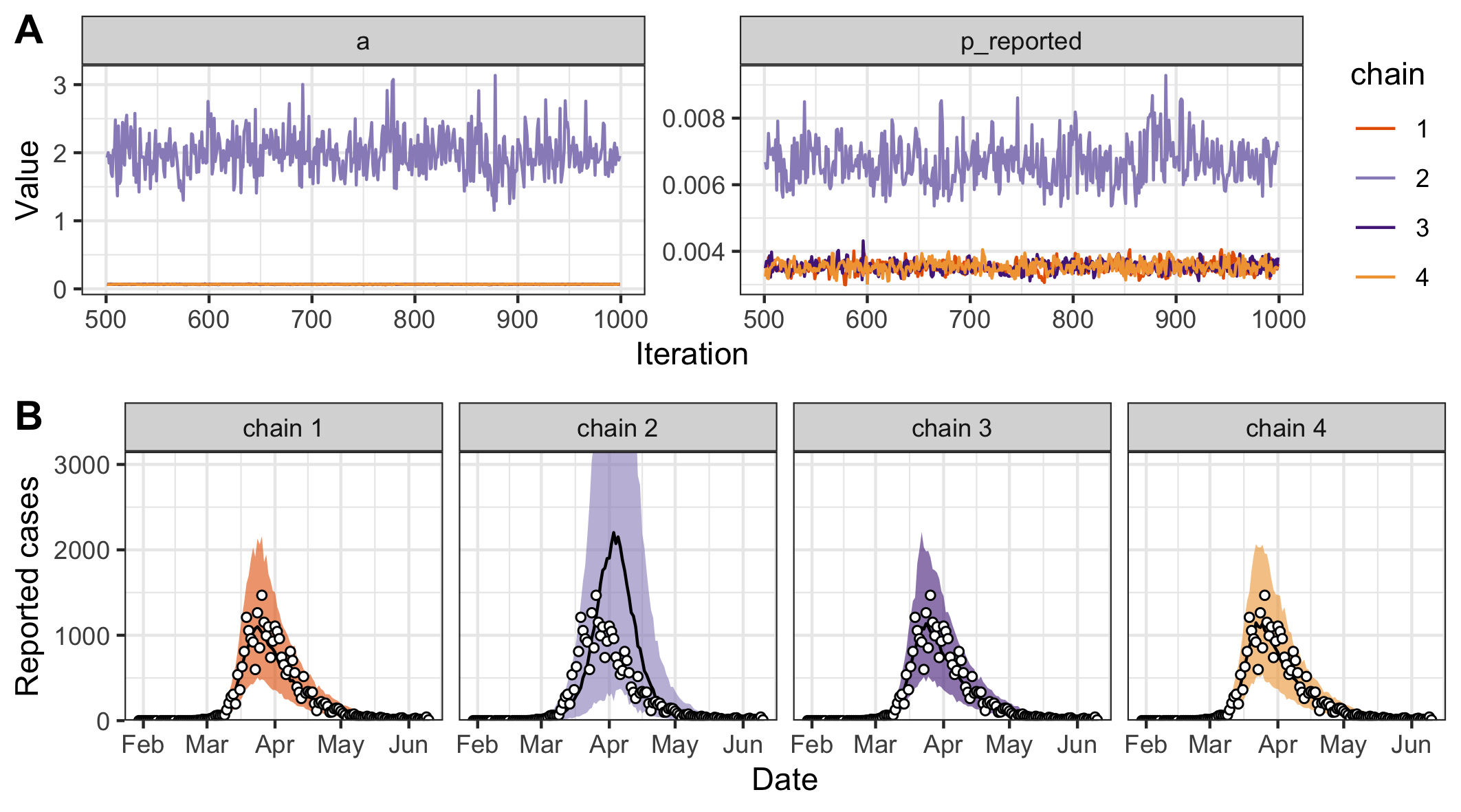}
%
%
%
	\caption{(A) Trace plot for two of the model parameters (the incubation rate $a$ and the reporting rate $p_r$ -- the other parameters are not shown) and (B) posterior predictive check of the number of reported cases for the SEIR model with underreporting (model iteration \#3) applied to data on the SARS-CoV-2 epidemic in Switzerland (white circles).}
     \label{fig:seir_diagnostics}
\end{figure}

Unfortunately, new obstacles appear when we try to fit this more complex model. \texttt{Stan} informs us that the chains are not mixing. To understand the issue, we try several diagnostics. Trace plots (Figure \ref{fig:seir_diagnostics}A) show that chains converge to two different modes when exploring the posterior distributions of the incubation rate $a$ and the reporting rate $p_r$. A chain-by-chain posterior predictive check (Figure \ref{fig:seir_diagnostics}B) shows that all four chains give better predictions than the previous models, but only chains 2 and 4 produce satisfying results. At this point, it is useful to look at the parameter spaces explored by different chains. In this case, we observe that chains 1 and 3 explore regions where the incubation rate $a$ is close to 2, so that the incubation time $1/a$ is very short, about 0.5 days.
However, we know from the literature that the average incubation time has been estimated around 5-6 days\cite{lauer2020incubation}.
This inconsistency suggests the incubation time cannot be inferred from the data alone and motivates incorporating more expert knowledge in the model in the form of an informative prior.
We reparametrize the incubation rate as its inverse, and set an informative prior on the incubation period centered around 6 days: $p(1/a) = \text{Normal}^+(6,1)$ (model iteration \#4).
Unfortunately, even after this correction, the chains still do not mix (not shown). 

We also notice that another parameter does not agree with domain knowledge: the probability that an individual infected with SARS-CoV-2 is reported as a case is very low for most chains, around 0.3-0.8\%. Such low values would imply that the entire population of Switzerland had been infected by July (30,994 reported cases over the period divided by 0.5\% gives more than 10 millions infected).
However, local serological studies \cite{stringhini2020seroprevalence} have shown that the cumulative number of infections in Geneva was not far from 10\% of the population by May 2020. This behavior of the current model can be explained. In this configuration of the model, transmission can only decrease due to acquired immunity (i.e. by lack of susceptible individuals). As the data show a decrease in reported cases, the model has to conclude that a large part of the population is immune at this point, and thus that the number of cases reported until then represents a very small proportion of the true number of infections. Given the data from the serological studies, the decrease in transmission must be due to something other than immunity.
Indeed on March 17, the Swiss government implemented lockdown measures to stop the spread of the disease.
Our model should therefore account for this effect.


\begin{figure}
    \centering
    \includegraphics[scale=0.2]{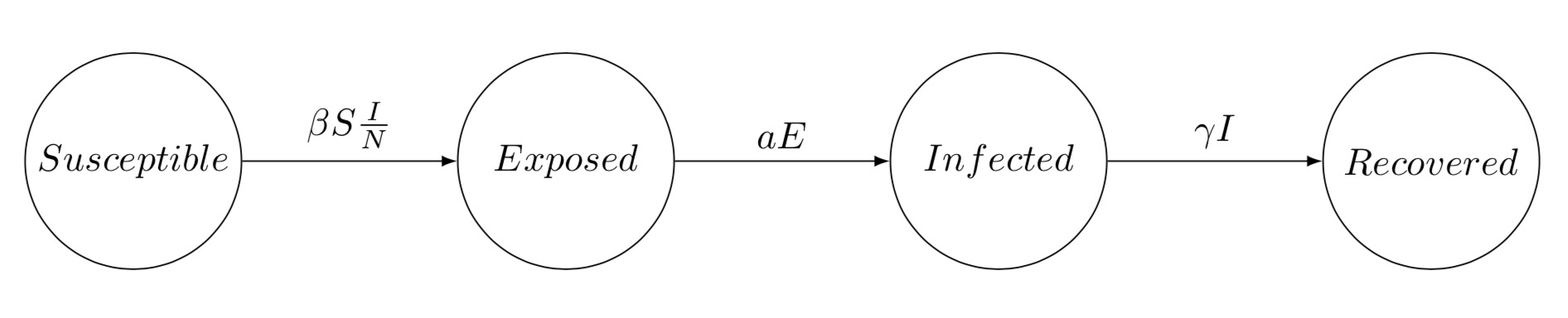}
    \caption{Diagram of a SEIR model.}
    \label{fig:seir}
\end{figure}

\subsection{Modeling control measures}
\label{sec:model_control}
We model the decrease in transmission after March 17 and the implementation of lockdown measures
\footnote{Note that we don't model explicitely the behavior changes from the population independent of the lockdown, so $f(t)$ could be interpreted as the combined effect of lockdown and independent behavior changes.} 
using a \textit{forcing function} with a logistic shape: 
$$\beta^*(t) = f(t) \beta,$$ 
with
$$f(t) = \eta + (1 - \eta)  \frac{1}{1 + \exp \left(\xi  (t - t_1 - \nu) \right)},$$
where $\eta$ is the decrease of transmission when control measures are fully in place, $\xi$ is the slope of the decrease, and $\nu$ is the delay (after the date of introduction of control measures, $t_1$) until the measures are 50\% effective (Figure \ref{fig:seir_survey_diagnostics}C). We add weakly-informative priors on the three parameters:
$p(\eta) = \mathrm{Beta}(2.5, 4)$ which means that we expect lockdown measures to reduce transmission, but not all the way to zero; $p(\xi) = \text{Uniform}(0.5, 1.5)$, which implies that the slope has to be positive but not too steep; and $p(\nu) = \text{Exponential}(0.2)$ which means that the delay before lockdown reaches half of its total efficiency should be between 0 and 20 days.

\begin{figure}
    \centering
    \includegraphics[scale=0.35]{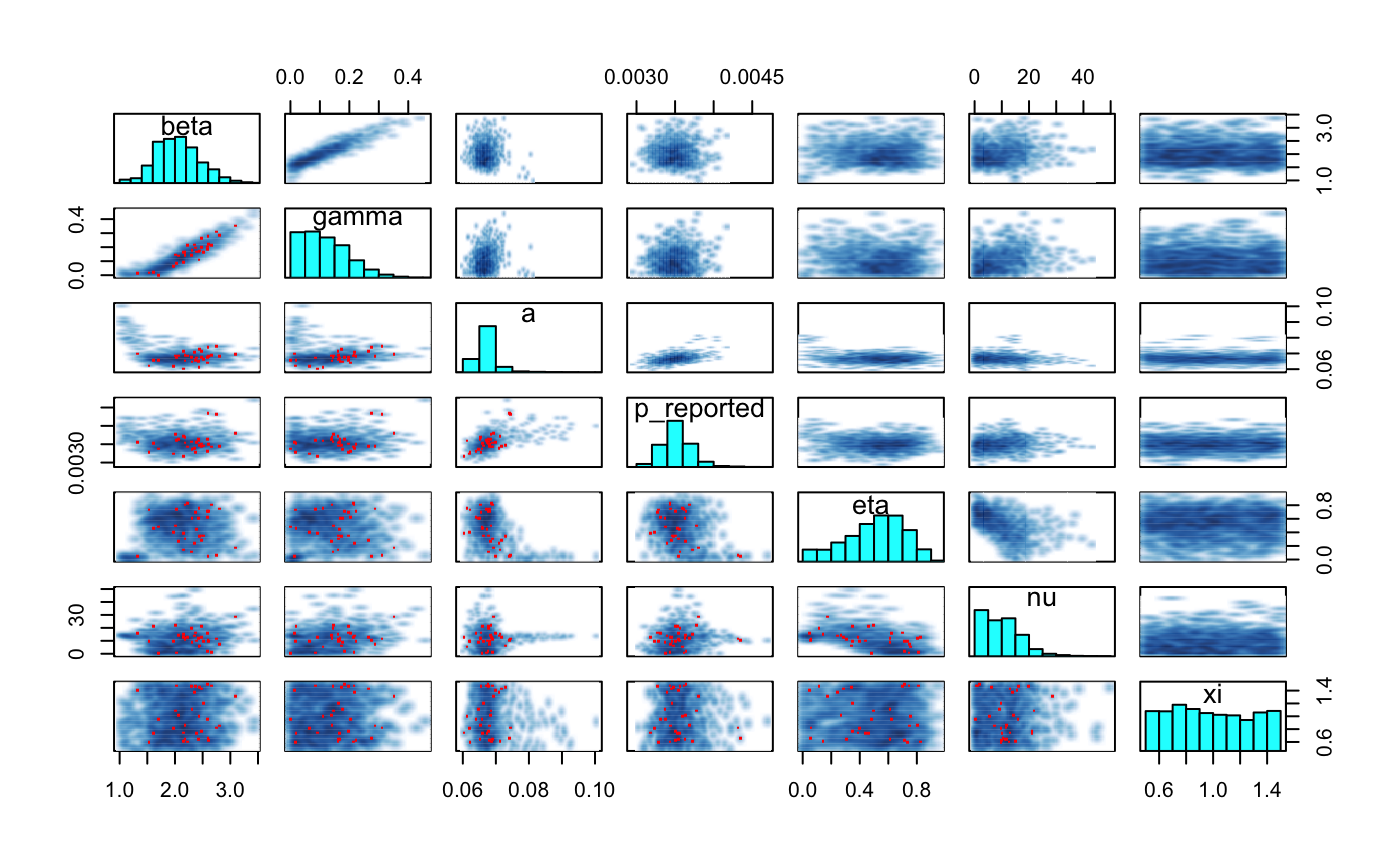}
    \caption{Pairs plot of all model parameters for the model incorporating control measures (model iteration \#5).}
    \label{fig:pairs}
\end{figure}

With this new formulation (model iteration \#5), the model should be able to describe the dynamics of SARS-CoV-2 transmission during the first wave of the pandemic.
However, our work is not finished yet, as the sampler appears to have trouble exploring the posterior distributions: \texttt{Stan} issues a warning about divergent transitions, which indicate the chains may not be exhaustively exploring the parameter space and are thus producing biased samples.
The other diagnostics are concerning: $\widehat R$ values go up to 1.19 and are above 1.01 for 6 of 8 model parameters, effective sample sizes are low, below 50 for 3 parameters.
Another concern is that the probability of a case being reported, $p_r$ remains very low, between 0.3 and 0.4\%. To understand what may be happening, we resort to another, more advanced diagnostic tool: the pairs plot (Figure \ref{fig:pairs}). The pairs plot shows samples across pairs of parameters and is useful to examine the geometry of the posterior density.
In this pair plot, divergent transitions, shown in red, do not concentrate in a particular zone of parameter space that could point towards a specific issue\cite{betancourt2013hier}.
We observe that the  posteriors are not well-defined bivariate bell curves but show some strong correlation (e.g., between $\beta$ and $\gamma$) and some irregular shapes (e.g. the spike between $a$ and $\nu$).
These points can be indicative of identifiability issues: several combinations of parameter values may result in the same observation.
Furthermore, we note that the marginal posterior distributions of $\eta$, $\nu$ and $\xi$, the three parameters controlling the effect of lockdown measures, are very similar to their prior distributions. 
This posterior behavior's is not surprising: when $p_r$ is that low, transmission can decrease substantially through immunity only, rendering the effect of lockdown measures non-identifiable, or at least degenerate\cite{betancourt2020identity}. 
We however believe that it is highly unlikely for $p_r$ to be very low and we could account for this using a more informative prior for $p_r$. 
In this case, we directly include serological survey results into the model, taking advantage of complementary data.

\subsection{Fitting data from a serological survey}
For simplicity, we only use data from week 5 of the survey. We ignore considerations about test sensitivity and specificity\cite{gelman2020sensitivity}, and consider that all recovered individuals will test positive to the serological test.
We also make the strong assumption that the results from this survey in Geneva are representative of the whole population of Switzerland. Tests from week 5 have taken place in Geneva from May 4 to May 7, and 83 out of 775 tested individuals were found to have antibodies against SARS-CoV-2. 
Given these assumptions, the probability of being detected by the survey is the proportion of individuals in the $R$ compartment at the time when the survey was conducted. 
We expand the sampling distribution defined in section \ref{sec:firstattempt}, multiplying the likelihood of data on reported cases by the likelihood of serological data:
$$
p(\mathcal Y' \mid \theta) = \text{Binomial}\left(\mathcal Y' \mid \frac{R(t_{\mathcal Y})}{N}, n_{\mathcal Y}\right)
$$
where $\mathcal Y'$ is the number of positive tests in the survey sample, $t_{\mathcal Y}$ is the time of the survey and $n_{\mathcal Y}$ is the sample size.
The implementation in code is straightforward (model iteration \#6).

\begin{figure}[h]
	\centering
	\includegraphics[width=.9\linewidth]{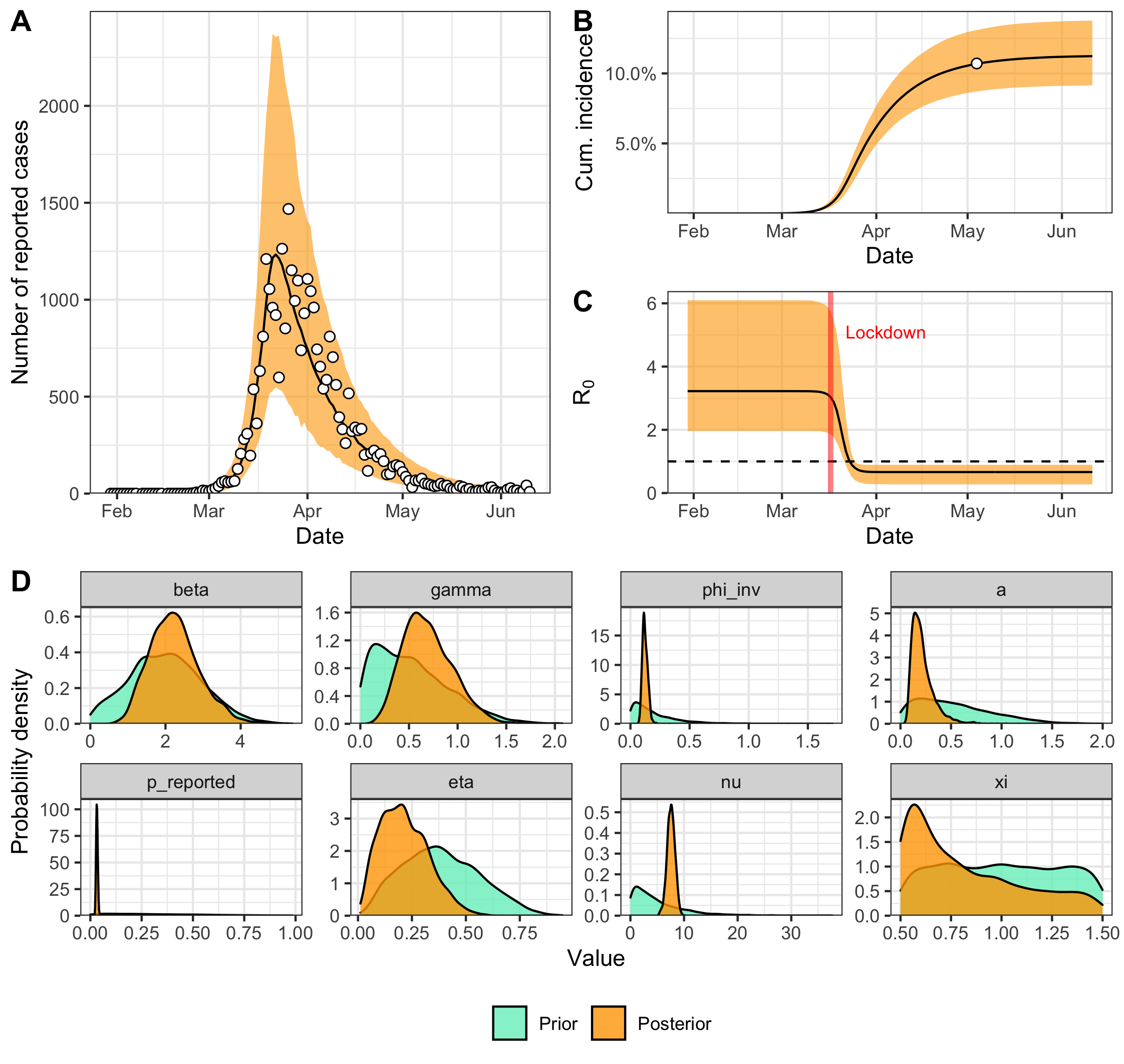}
	\caption{(A) Posterior predictive check of the number of reported cases and (B) of the cumulative incidence for the SEIR model including the effect of control measures and fitted to both reported cases and seroprevalence data (model iteration \#7; white circles show data on reported cases in panel A and seroprevalence data in panel B). (C) Posterior distribution of the forcing function $f$ that models the reduction in transmission after the introduction of lockdown measures. (D) Prior and posterior distributions of the parameters of model iteration \#7.}
	\label{fig:seir_survey_diagnostics}
\end{figure}

Running this new iteration of the model, \texttt{Stan} issues a warning that the tree depth is often exceeded.
This means that the sampler had to choose a step size small enough to explore some part of the posterior, but that this step size is too small for exploring another part of the posterior efficiently, slowing down the sampling process. This is not too worrying, given that the other diagnostics are good.
We can increase the maximum tree depth and try again (model iteration \#7).
Finally this time, no warning is issued, all diagnostics are good, and the prior and posterior predictive checks confirm the reliability of the inference and the relative adequacy of the model (Figure \ref{fig:seir_survey_diagnostics}A-B).
We can go forward with the application of the model and discuss the results -- always in the context of the model assumptions and priors.
For instance, we might want to report that we estimate the basic reproduction number $\mathcal R_0$ to be 2.7 (95\% credible interval 1.9-5.2) and the reporting rate 3.3\% (2.7-4.1). We can also focus on the effects of lockdown measures, interpreting $1-\eta$ as a relative reduction of 73\% (53-92) in transmissibility after lockdown measures are fully efficient, or interpreting $\nu$ as an implementation delay of 7.5 days (6.2-8.9) before lockdown measures are half-efficient.

We finally obtained a decent model, but note that its aim is only didactic and should not be directly used to inform policy. 
The first functioning model is not the end of the road, and a lot of features should be considered before claiming to have obtained a good depiction of the Swiss epidemic.
For instance, one could account for the sensitivity and specificity of tests when fitting seroprevalence data, improve the inference by including data on testing, hospitalisations and deaths, relax some of the strong assumptions that were made, and stratify by age, location or some other characteristic. 
This would result in a large number of competing models that can be organised in a \textit{network} depending on what features are included\cite{gelman2020bayesian}.
Model comparison tools become invaluable in this context.
Although it is out of the scope of this tutorial, we recommend using \textit{leave-one-out cross-validation}\cite{loo2017}, or, even more adapted for time series, \textit{leave-future-out cross-validation} \cite{burkner2020approximate}.

 
%

%

%% file: table_covid_extension.tex
\begin{table}[]
\centering
\begin{adjustbox}{width=\textwidth}
\begin{tabular}{|c|l|l|l|l|l|l|}
\hline
Iteration &
  Model &
  Data &
  Stan diagnostics &
  Other issues &
  Interpretation / Comments &
  Proposed improvement \\ \hline
\#1 &
  Basic SIR model &
  Reported cases &
  \begin{tabular}[c]{@{}l@{}}$\hat{R} >> 1$,\\ max\_treedepth\\exceeded\end{tabular} &
  Far-off predictions &
  Misspecified model &
  \begin{tabular}[c]{@{}l@{}} Add a reporting rate \\parameter $p_r$ \end{tabular}
   \\ \hline
\#2 &
  \begin{tabular}[c]{@{}l@{}}SIR + \\underreporting\end{tabular} &
  Reported cases &
  None &
  Skewed prediction &
  Misspecified model &
  \begin{tabular}[c]{@{}l@{}} Add incubation rate\end{tabular} \\ \hline
\#3 &
 \begin{tabular}[c]{@{}l@{}}SEIR + \\underreporting + \\ incubation rate\end{tabular} &
  Reported cases &
  $\hat{R} > 1.01$ &
  \begin{tabular}[c]{@{}l@{}}1 out of 4 chains\\ with skewed predictions\\ and unrealistic incubation time\end{tabular} &
  Degeneracy due to weak priors &
   \begin{tabular}[c]{@{}l@{}} More informative prior \\on incubation time\end{tabular} \\ \hline
\#4 &
  \begin{tabular}[c]{@{}l@{}}SEIR +\\ underreporting + \\varying initial infections +\\ informative prior on incubation time\end{tabular} &
  Reported cases &
  $\hat{R} > 1.01$ &
   \begin{tabular}[c]{@{}l@{}}Unrealistic values of $p_r$ \\for some chains\end{tabular} &
  \begin{tabular}[c]{@{}l@{}}The model can only make\\ infections  decrease \\if $p_r$ is very low\end{tabular} &
  Model control measures \\ \hline
\#5 &
  \begin{tabular}[c]{@{}l@{}}SEIR + \\underreporting + \\varying initial infections +\\ model control measures\end{tabular} &
  Reported cases &
  A few divergences &
  \begin{tabular}[c]{@{}l@{}}$p_r$ is unrealistic,\\ degenerate values for\\ logistic parameters\end{tabular} &
  \begin{tabular}[c]{@{}l@{}}Degeneracy between the two ways \\ to make infections decrease: \\ immunity and control measures\end{tabular} &
  \begin{tabular}[c]{@{}l@{}}  Inform $p_r$ by including \\ serological data\end{tabular}\\ \hline
\#6 &
  \begin{tabular}[c]{@{}l@{}}SEIR + \\underreporting +\\ varying initial infections +\\ model control measures\end{tabular} &
  \begin{tabular}[c]{@{}l@{}} Reported cases
  \\ + serology 
  \end{tabular} &
  \begin{tabular}[c]{@{}l@{}}max\_treedepth \\exceeded for most \\iterations,\\  a few divergences\end{tabular} &
  None &
  Unclear &
  \begin{tabular}[c]{@{}l@{}}  Increase the maximum \\ tree depth \end{tabular} \\ \hline
\#7 &
  \begin{tabular}[c]{@{}l@{}}SEIR + \\underreporting + \\varying initial infections +\\ model control measures + \\ increased max tree depth\end{tabular} &
  \begin{tabular}[c]{@{}l@{}} Reported cases\\ 
   + serology \end{tabular} &
  None &
  None &
  The model seems decent! &
  None \\ \hline
\end{tabular}%
\end{adjustbox}
\caption{Summary of the Bayesian workflow used for iteratively building a model of SARS-CoV-2 transmission in Switzerland.}
\label{tab:table_covid}

\end{table}

%% file: conclusion.tex
Modeling is an iterative process.
We rarely start with a good model.
Rather we must build our way to a good model, starting from a baseline, and revising our models as we uncover shortcomings.
The Bayesian modeling workflow offers a perspective, which goes beyond fitting a polished model or using inference algorithms in an idealized context; it encompasses failing cases, and techniques to diagnose and learn from these failures.
In order to navigate the workflow, we must reason about the modeled phenomenon, our inference algorithms, our computational implementation, and how all these elements interact with one another.
This means leveraging domain, statistical, and computational expertise.

We show the advantages that can be obtained from this Bayesian workflow in epidemiology.
\texttt{Stan} is an adequate tool for building, fitting, and criticizing ODE-based models, as we demonstrate on an influenza model and a more sophisticated SARS-CoV-2 model.
For the latter, several ingredients are required to build an adequate model.
First and foremost, using an appropriate epidemiological model for a given disease is key.
Secondly, in order to improve our estimates of the parameters and overcome issues of identifiability, we need to incorporate information by using well-motivated priors and combining data from multiple sources.
Finally, we must tune our inference algorithms in order to accurately probe the posterior distribution.
We discover these ingredients by gradually building our model and deploying a broad range of diagnostics.
One important practical point is that we need computationally efficient implementations of our code to get reasonably fast inference for the final model and also to quickly troubleshoot failing models.

As a final thought, we note that the modes of failures for models we develop along the way improve our understanding of the final model.
Sharing these early models can therefore make our work more transparent and improve scientific communication.

\section*{Code}
\label{sec:code}

The code to run the examples in this article can be found at \url{https://github.com/charlesm93/disease_transmission_workflow}.

